\documentclass[pre,aps,twocolumn,showpacs,floatfix]{revtex4-1}
\usepackage{amssymb}

\usepackage{graphicx}  
\usepackage{dcolumn}   
\usepackage{bm}        
\usepackage{amsmath}
\usepackage{slashbox}
\usepackage{float}

\begin{document}

\title{Properties of long quantum walks in one and two dimensions}

\author{Luo Hao,$^1$ and Peng Xue$^{2,3}$}

\affiliation{$^1$Department of Physics, Renmin University of China, Beijing
100872, China \\
$^2$Department of Physics, Southeast University, Nanjing 211189, China \\
$^3$State Key Laboratory of Precision Spectroscopy, East China Normal
University, Shanghai 200062, China}

\date{\today}


\begin{abstract}

The quantum walk (QW) is the term given to a family of algorithms
governing the evolution of a discrete quantum system and as such has a
founding role in the study of quantum computation. We contribute to the
investigation of QW phenomena by performing a detailed numerical study
of discrete-time quantum walks. In one dimension (1D), we compute the
structure of the probability distribution, which is not a smooth curve
but shows oscillatory features on all length scales. By analyzing walks
up to $N$ = 1000000 steps, we discuss the scaling characteristics and
limiting forms of the QW in both real and Fourier space. In 2D, with a
view to ready experimental realization, we consider two types of QW,
one based on a four-faced coin and the other on sequential flipping of
a single two-faced coin. Both QWs may be generated using two two-faced coins, which in the first case
are completely unentangled and in the second are maximally entangled.
We draw on our 1D results to characterize the properties of both walks,
demonstrating maximal speed-up and emerging semi-classical behavior in
the maximally entangled QW.

\end{abstract}

\maketitle

\section{\label{sec1} Introduction}

Quantum walk(QW) is an extension of the classical random walk, in which the walker
is controlled by a quantum mechanical variable \cite{a2,PhysRevA.48.1687}.
The quantum interference from different walk paths brings about QW novel and
interesting features. An motivation for work on QW is developing new quantum algorithms \cite{PhysRevA.67.052307,r1,r2},
aiming at an exponential speed-up compared to classical algorithms for a certain classes of problem \cite{a23,k1,am1}.

On the physical side, QW is a valuable model for studying evolution process
from simple quantum protocol: discrete-time QW \cite{a11} and continuous-time QW
in one dimension (1D) \cite{a24}. Deep studies involve multiple coins
\cite{a4}, multiple walkers \cite{a6}, multiple dimensions \cite{a3}, and multiple coins with different
degrees of entanglement \cite{px1,a8,a9}. A recent resurgence of interest in the QW
has come about to a significant extent because experimental capability has
caught up with many of these theoretical proposals, particularly to create
walks with complex coin protocols and in two phase-space dimensions
\cite{Do:05,PhysRevLett.104.050502,e1,PhysRevA.81.052322,a33,PhysRevLett.103.090504,PhysRevLett.104.100503,a32,e2,Schreiber06042012}.
However, all of these QWs implementation are restricted to rather small
numbers of steps, and some have inherent limits to the size of their
phase space.

Despite the large volume of work performed on the QW, we have not been
able to find any that address the properties of the system at long evolution
times, meaning at large values of the step number $N$. As the challenges
in quantum computing move toward the large scale, it would appear that there
is a need to understand the behavior of quantum algorithms in the large-$N$
regime. The aim of this paper is to analyze QWs at large $N$ to establish
their properties, both universal and specific, and thus effectively to gauge
the flow and concentration of information in one particular set of algorithms.
We will consider in detail the 1D QW, where we have performed calculations
up to $N = 1000000$ steps, to establish the scaling characteristics,
spatial information content, universality, and limiting form of the
probability distribution. We will then turn to different types of possible
QW in 2D, where we use our 1D knowledge both directly and to benchmark
the additional forms of behavior that emerge, particularly entanglement,
accelerated diffusion, and semi-classical limiting distributions.

In this paper we perform a detailed analysis of the discrete-time 1D QW and two
fundamental 2D QWs. By calculating probabilities for large numbers $N$ of
steps, we investigate the destructive interference, the scaling properties,
the frequency content, and the combination of QWs. The structure of the
article is as follows. In Sec.~II we review the classical random walk and
a simple, symmetrical model for understanding the discrete-time QW in
1D, including its analytical solution. Section III presents our numerical
results for 1D QWs up to large $N$ and their analysis in both real and
Fourier space. In Sec.~IV we discuss two typical 2D QWs as a theoretical basis of Sec.~V.
 Borrowing from the understanding developed in 1D, in Sec.~V we
provide the complete numerical characterization of these two 2D QWs.
Section VI provides a brief summary.

\section{\label{sec2} One-dimensional Random Walks}

\subsection{\label{sec2a} Classical Random Walk}

\begin{figure}[t]
\centering
\includegraphics[width=0.48\textwidth]{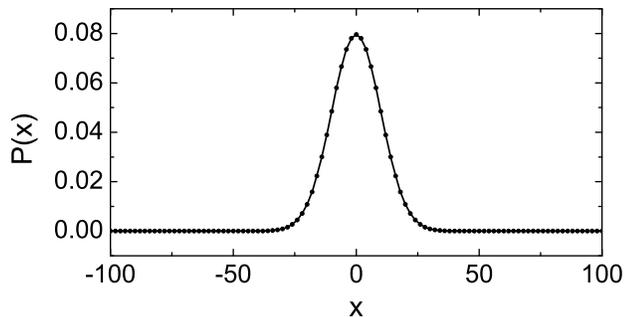}
\caption{Probability distribution for a 1D classical random walk with
$N = 100$.}
\label{fig:1dcrw}
\end{figure}

A walker standing at the origin of a line flips an unbiased coin and steps
to the right if a head comes up or to the left if the result is a tail. After
many flips, and taking a fixed number of steps based the coin state, the
location of the walker is an unknown, random position, but the probability
distribution of this position is a definite statistical quantity. After $N$
steps, the probability of the walker being at position $x$ is
\begin{equation}
P_N(x) = \frac{1}{2^N} C_N^{\frac{N-x}{2}} \;\; = \;\; \left( \frac{1}{2} \right)^N
\frac{N!}{\left( \frac{N+x}{2} \right) \! \left( \frac{N-x}{2} \right)!},
\label{eq01}
\end{equation}
i.e.~this process, the discrete-time 1D classical random walk, follows a
binomial distribution. The walker's position $x$ may only be an even (odd)
integer when $N$ is even (odd). In the continuum limit, which is approached
for sufficiently large $N$, the distribution is Gaussian (Fig.~\ref{fig:1dcrw}).
Its standard deviation, which represents a mean propagation distance in a
sample of many walkers, is $\sigma = \sqrt{N}$.

\subsection{\label{sec2b} Quantum Walks}

\begin{figure}[t]
\centering
\includegraphics[width=0.48\textwidth]{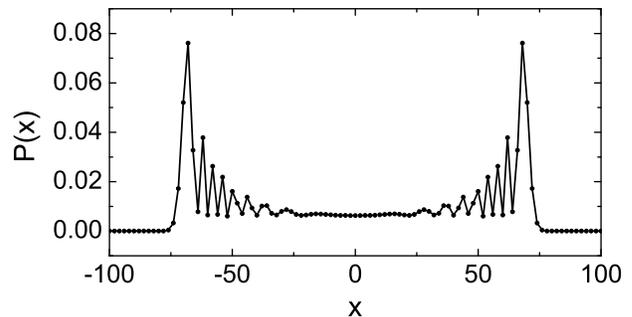}
\caption{Probability distribution for a 1D quantum walk with
$N = 100$.}
\label{fig:1dqrw}
\end{figure}

A QW refers to a random walk effected using a quantum device, and as such
requiring a quantum mechanical description.Following the classical formulation of a coin and a walker, the QW differs fundamentally from its classical counterpart in that coin and walker are ``entangled'' in the same quantum
``particle''. The essential consequence of this entanglement is that the propagation
of the quantum walker depends not on its probability (Sec.~IIA) but on its
amplitude. This amplitude is subject to quantum mechanical interference,
which can be constructive or destructive, leading to completely unconventional
forms of probability distribution (Fig.~\ref{fig:1dqrw}).

The numerical results in Sec.\ref{sec3} based on following calculation.
While there are many ways of describing such a walk (Ref.~\cite{a2} for a review),
here we introduce only the case of a discrete time symmetric QW on an open line.
With a view to later application, we denote the two internal ``coin''
states of the walker, or particle, as $\uparrow$ and $\downarrow$.
In a classical walk the coin states are completely separate
($\uparrow$ or $\downarrow$ with probabilities 0 or 1), whereas a quantum
coin can occupy any superposition state $a |\uparrow \rangle + b |
\downarrow \rangle$. Thus the quantum system is described by the wave
function
\begin{equation}
|\psi_N \rangle = \sum\limits_{i \, = -N}^{N} (a_i |\downarrow \rangle
 + b_i |\uparrow \rangle) |i \rangle,
\label{eqrwp}
\end{equation}
where $N$ is number of steps in the walk, $i$ is the position index, and
$|i \rangle$ the corresponding state.

The QW is described by the
evolution of this wave function under the quantum operation for successive
steps. The probability distribution for finding the walker at position $i$
(state $|i \rangle$) is found from the trace over the coin states to
be $P_i = a_i^2 + b_i^2$, with $\sum_i P_i = 1$.

The evolution has two substeps: (1) A rotation in the coin space, represented
by a unitary operator $U$, which is often taken as the Hadamard transformation
\begin{equation}
U_H = \frac{1}{\sqrt{2}} \left( \!\! \begin{array}{cc} 1 & 1 \\ 1 & -1
\end{array} \!\! \right) \! .
\label{eho}
\end{equation}
(2) A coin-dependent translation of the walker, described by
the shift operator
\begin{equation}
S = |\uparrow \rangle \langle \uparrow |\otimes \sum\limits_i
|i+1 \rangle \langle i |\, + |\downarrow \rangle \langle \downarrow |
\otimes \sum\limits_i |i-1 \rangle \langle i|,
\label{eso}
\end{equation}

The coefficients $a_0$ and $b_0$ of the initial coin state are arbitrary,
which has a profound influence on the symmetry of the probability distribution
of the QW \cite{a2}. To obtain a distribution symmetric under the application
of the Hadamard operator, the initial coin state is written as
\begin{equation}
|\psi_0 \rangle = {\textstyle \frac{1}{\sqrt{2}}} (|\uparrow \rangle +
i |\downarrow \rangle) \otimes |0 \rangle,
\label{002}
\end{equation}
A full step of the QW is
\begin{equation}\label{esh}
|\psi_1 \rangle = S U_H |\psi_0 \rangle = \frac{1+i}{2} |\uparrow
\rangle |1 \rangle + \frac{1-i}{2} |\downarrow \rangle |-1 \rangle
\end{equation}

The probability distribution for a QW of $N = 100$ steps is shown in
Fig.~\ref{fig:1dqrw}. It is manifestly completely different from the
classical random walk (Fig.~\ref{fig:1dcrw}), with clear maxima at values of
$i$ close to $\pm \, 0.7 N$ (revealed at larger $N$ to be $\pm \, N/\sqrt{2})$.
The probability for $i = 0$ is close to 0, indicating that the origin of this
counter-intuitive behavior is an almost complete destructive interference
among the paths of the quantum walker returning to the origin. The standard
deviation of the probability distribution in this QW is $\sigma \propto N$,
indicating a linear spreading rate, which is one of the most important
attributes of this evolution algorithm [Eq.~(\ref{esh})]. We present a
quantitative analysis of the properties of the QRW in Sec.~III. The QW
evolution process acts to propagate the quantum mechanical amplitudes,
preserving the complete information content of the internal states.

There are another two kinds of analytical descriptions
do benefit to understand the evolution of QW.
The difference between the classical and quantum walks can be understood
from the non-commuting nature of the (matrix) quantum operators and its
consequences for the interference of different walker paths. The Hadamard
operator can be decomposed \cite{a10} as $U_H = P + Q$ with $$P = \frac{1}
{\sqrt{2}} \left( \! \begin{array}{cc} 1 & 1 \\ 0 & 0 \end{array} \! \right)
\; {\rm and} \;\; Q = \frac{1}{\sqrt{2}} \left( \! \begin{array}{cc} 0 & 0
\\ 1 & -1 \end{array} \! \right) \! .$$It is easy to see that $P$ determines
motion of the walker to the right and $Q$ to the left. Evolution under the
QW for $N$ steps is represented as $U_H^N = (P + Q)^N$. In the classical
random walk one has $1^N = (p + q)^N$, where $p = q = \frac{1}{2}$ are real numbers representing probabilities.

One may conclude that the non-commutativity of the quantum operators determining different paths
to the same walker position, encodes the interference of amplitudes leading to the entirely unconventional
quantum phenomena reflected in Figs.~\ref{fig:1dcrw} and \ref{fig:1dqrw}. We expand this method to study a novel 2D-QW behaviour in Sec.\ref{sec5b}.

We pay attention to another analytical solution\cite{a11} in order to introduce the concept of Fourier
analysis. The evolution operator has a more concise form in $k$-space (Fourier
space) than in $x$-space (real space), so the initial wave function may be
transformed to and evolved in $k$-space. By inverse Fourier transformation, the real-space wave functions for the
$|\uparrow \rangle$ and $|\downarrow \rangle$ internal states are

\begin{widetext}
\begin{equation}
\begin{split}
\psi_\uparrow (x,t) = \frac{1}{2\pi} \int_{-\pi}^{\pi} \!\!\! dk e^{-ikx} &
\frac{e^{-i\omega_kt} (\sqrt{1 + \cos^2k} + \cos k + ie^{-ik}) + e^{i(\omega_k - \pi)t}
(\sqrt{1 + \cos^2k} - \cos k - ie^{-ik})}{2\sqrt{2}\sqrt{1 + \cos^2k}}, \\
\psi_\downarrow (x,t) = \frac{1}{2\pi} \int_{-\pi}^{\pi} \!\!\! dk e^{-ikx} &
\frac{ i [e^{-i\omega_kt} (\sqrt{1 + \cos^2k} - \cos k - i e^{ik}) + e^{i(\omega_k
 - \pi)t} (\sqrt{1 + \cos^2k} + \cos k + i e^{ik})]}{2\sqrt{2}\sqrt{1 +
\cos^2k}}, \\
\end{split}
\label{f01}
\end{equation}
\end{widetext}
and finally the probability of finding the walker at a given position $x$
after a walk of $t$ steps is given by
\begin{equation}
P(x,t) = |\psi_\uparrow (x,t)|^2 + |\psi_\downarrow (x,t)|^2.
\label{f02}
\end{equation}
Eqs.~(\ref{f01}) and (\ref{f02}) contain no dynamics,
because the term $e^{-i\omega_kt}$ provides no more than a compact notation for
the combination of $N$ with $k$, the spatial Fourier variable conjugate to
the actual walker displacement $x$.

\begin{figure}[!htpb] \center
\includegraphics[width=0.5\textwidth]{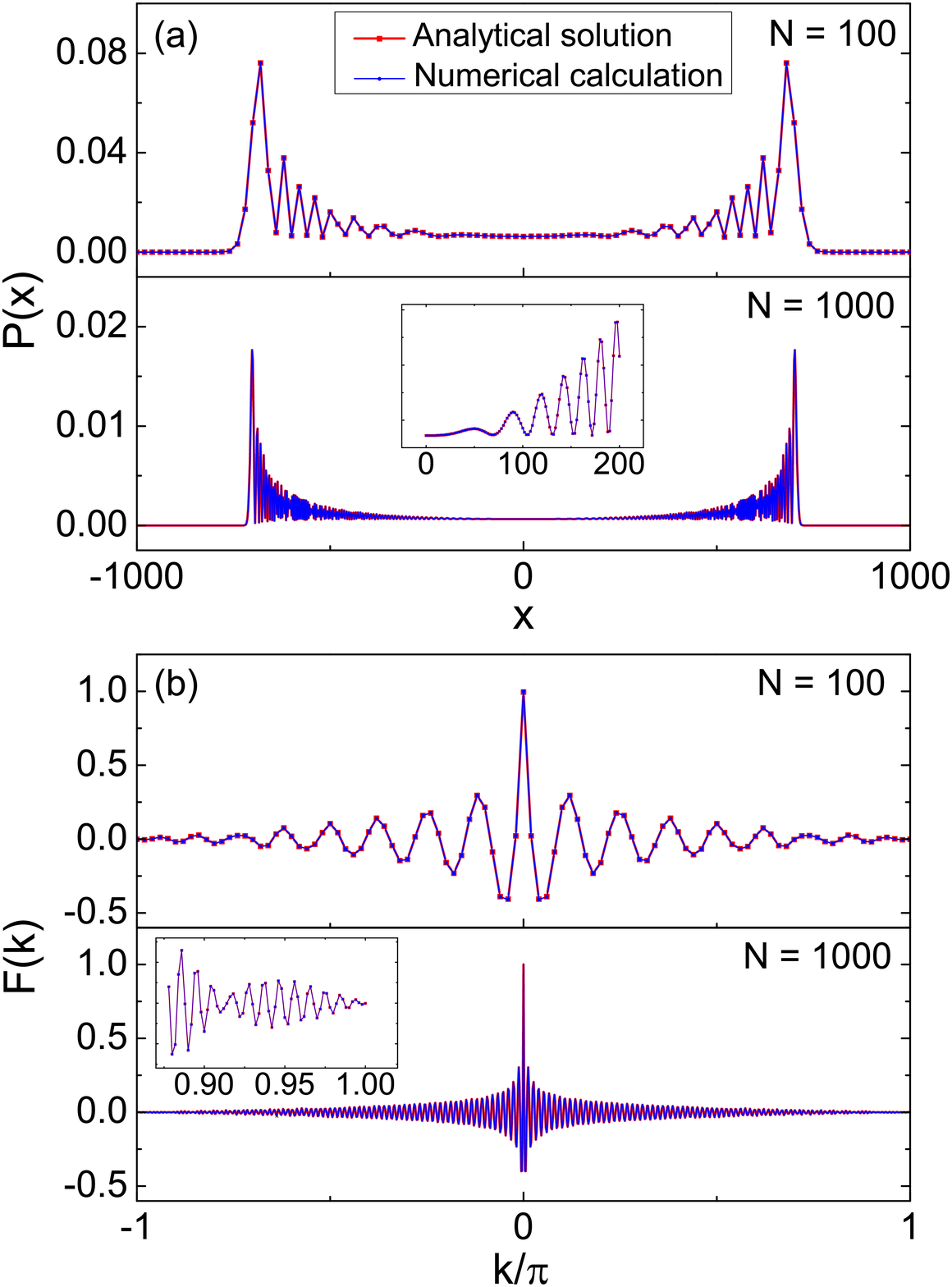}
\caption{(color online) Probability distribution for the 1D QW, comparing
the exact expression of Eqs.~(\ref{f01}) and (\ref{f02}), shown in red,
with numerical calculations based on Eqs.~(\ref{002}) and (\ref{esh})
shown in blue, for walks of $N = 100$ and 1000 steps. (a)
Probability distribution $P(x)$ in real space. (b) Fourier components $F(k)$
of $P(x)$. Insets for $N = 1000$ show (a) the probability oscillations near
$x = 0$ and (b) beats in the Fourier envelope near $k = \pi$ (see text). }
\label{num-ana}
\end{figure}

A numerical calculation of the exact analytical solution, contained in
Eqs.~(\ref{f01}) and (\ref{f02}), is shown as the red curves and points in
Fig.~\ref{num-ana}, where we compute the probability distribution in both
real space [Fig.~\ref{num-ana}(a)] and Fourier space [Fig.~\ref{num-ana}(b)]
for $N = 100$ and 1000. The blue curves show our numerical calculations
based on method in Eqs.~(\ref{esh})
with the initial state specified in Eq.~(\ref{002}). The results are
identical up to a relative error of $10^{-5}$ caused by the numerical
integration of Eq.~(\ref{f01}). We comment in detail on the forms of
these distributions in Sec.~III.

The importance of the values $\pm \, N/\sqrt{2}$, noted in Sec.~IIB, is
clearly evident in Fig.~\ref{num-ana}(a), and it was deduced in Ref.~\cite{a25}
that the limiting distribution is concentrated in the interval $\left[
-\frac{N}{\sqrt{2}}, \frac{N}{\sqrt{2}} \right]$ as $N \rightarrow
\infty$. We will qualify this statement in Sec.~IIIA. Although the analytical
solution \cite{a11,a25} gives the exact probabilities for any position and
number of steps, in fact the expressions in Eq.~(\ref{f01}), which require
numerical integration over complex quantities, are not easy to compute when
$N$ becomes large. In this regime, direct numerical calculation of the
probability distribution is more straightforward, and we use this approach
in Sec.~III to reveal the properties and structure of the 1D QW at large $N$.

\section{\label{sec3} Numerical Studies of the 1D QW}

We now consider in detail the properties of the 1D QW introduced in
Sec.~II. The probability distribution obtained from this quantum
algorithm has a number of very striking properties in both real and
Fourier space (Fig.~\ref{num-ana}). Among them are the clear importance
of $\pm \, N/\sqrt{2}$ noted above, the twin-peaked ``envelope function''
of the $P(x)$ distributions with its remarkable zone of destructive
interference around $x = 0$, the width of these peaks, and the rapid
oscillation of the functions (both $P(x)$ and $F(k)$) at high spatial
frequencies within the envelope. We clarify immediately that this
oscillatory behavior is ``real'' in the sense that Figs.~\ref{fig:1dqrw}
and \ref{num-ana} show only the probabilities at even steps 0, $\pm \, 2$,
$\pm \, 4$, $\dots$, with the zero-probability odd steps not shown; the
QRW contains additional oscillations in space between the ``microscale''
of individual steps and the ``macroscale'' of the walk length.

\begin{figure}[t] \center
\includegraphics[width=0.5\textwidth]{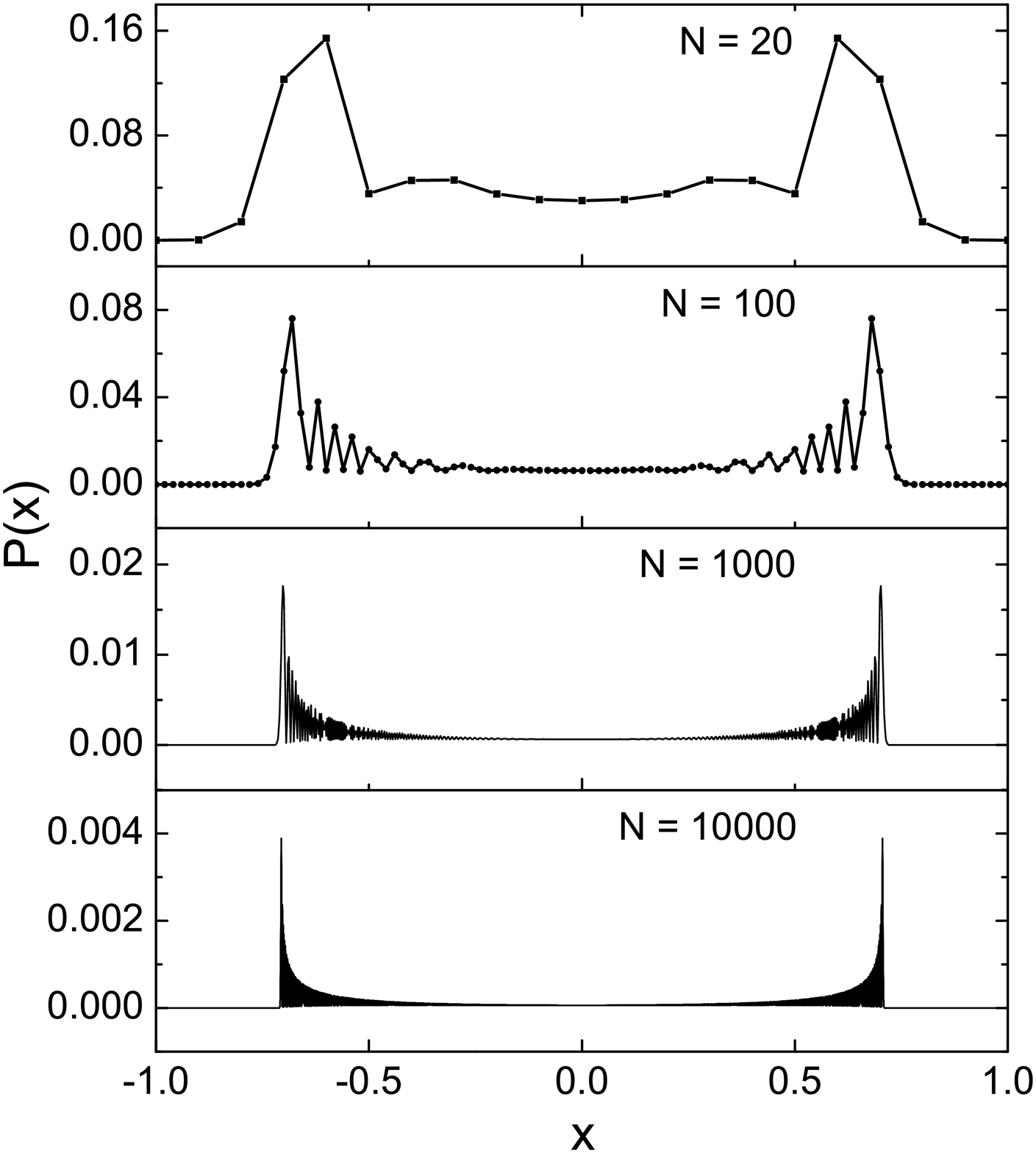}
\caption{Numerical results for the 1D QW with different numbers $N$ of
steps, scaled to the same horizontal axis. Probabilities are shown only for
even values of the site position $x$, with zero values on odd sites excluded
from the curves.}
\label{1001e}
\end{figure}

\subsection{\label{sec2d2} $N/\sqrt{2}$ Property}

We begin by analyzing the most obvious feature of the QW, which is the
tendency for the probability distribution to peak around 0.7$N$. From
Sec.~IIC it is clear that the factor $1/\sqrt{2}$ plays an important
role in the analytic solution and in physical terms it would appear to
mark the crossover in behavior from a low probability arising due to
destructive interference to a low probability arising simply from the
extremely low likelihood of having more than 85\% of the steps of an
unbiased random walk be in the same direction. In Fig.~\ref{1001e} we
show numerical results for the probability distributions of 1D QWs
with four different values of $N$, with the position axis normalized
by $N$. As $N$ increases, the distributions exhibit both increasingly
oscillatory behavior, which we analyze in Sec.~IIIB, and a peaking of
the envelope function, which becomes sharper as it converges towards
a maximum probability close to step $N/\sqrt{2}$ (Sec.~IIIC).

\begin{figure}[t] \center
\includegraphics[width=0.45\textwidth]{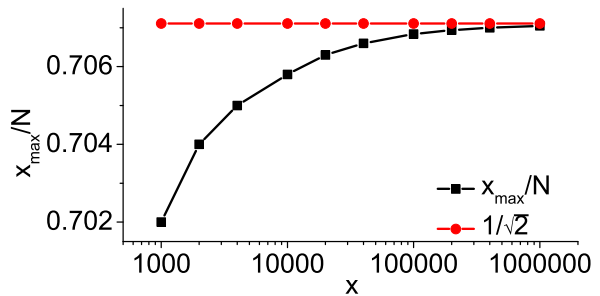}
\caption{(color online) Scaling of the position $x_{\rm max}$ for
maximum probability with total step number $N$ in the 1D QW.}
\label{1001t}
\end{figure}

\begin{figure}[t] \center
\includegraphics[width=0.45\textwidth]{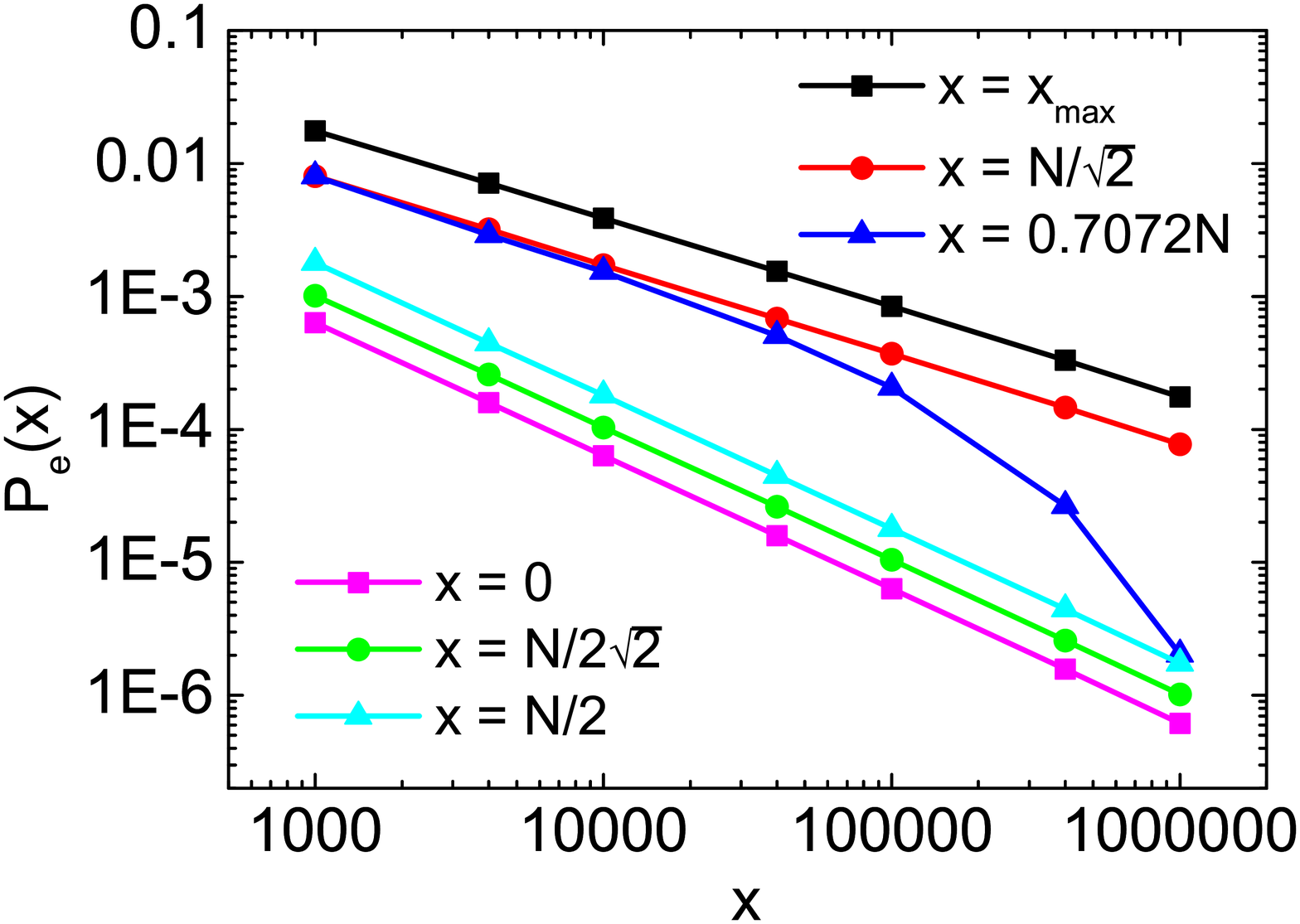}
\caption{Representative probability data from different points on the
distribution, $x = 0$, $x = N/2\sqrt{2}$, $x = N/2$, $x = x_{\rm max}$,
$x = N/\sqrt{2}$, and $x = 0.7072N$, for 1D QWs of step lengths from
$N = 1000$ to $N = 1000000$. Where relevant, all data are taken from
the upper envelope of the distribution.}
\label{1004t}
\end{figure}

This convergence in shown in more detail for values up to $N = 1000000$
in Fig.~\ref{1001t}. The macroscopic feature is indeed a convergence
towards $N/\sqrt{2}$. Further, on a relative scale the distribution
appears to tend towards $\delta$-functions centered at $\pm N/\sqrt{2}$.
However, we caution that this is not the complete story and we consider the
asymmetric envelope shape in Sec.~IIIC. The point of maximum probability is
in fact $x_{\rm max} = 70684$ for $N = 100000$ and $x_{\rm max} = 707050$
for $N = 1000000$, while the exact value of $1/\sqrt{2}$ is 0.707106.
Thus in fact there are still more than 25 even steps of finite probability
separating $x_{\rm max}$ from $N/\sqrt{2}$ for $N = 1000000$ [see
Figs.~\ref{qob}(i) and \ref{qob}(j)]. The probability at $x = N/\sqrt{2}$,
shown in Fig.~\ref{1004t}, is always close to one half of $P(x_{\rm max})$,
and this point marks the approximate crossover where $P(x,N)$ changes from
algebraic to exponential decay with $N$ (below). There are always points
of finite $P(x)$ beyond $x = N/\sqrt{2}$ and the more exact statement of the
result of Ref.~\cite{a25} is that the normalized support converges to the
interval $\left[ -\frac{1}{\sqrt{2}}, \frac{1}{\sqrt{2}} \right]$ as $N
\rightarrow \infty$. The probability of a walker passing beyond this interval
vanishes more rapidly than interference effects can cause it to grow, and the
net consequence of the destructive interference between paths is to ``pile
up'' the probability close to (but mostly below) the limits of the interval.

Further insight into the nature of the (upper) envelope function may be
obtained by considering the probabilities at different representative
positions on the normalized $x$-axis, as shown in Fig.~\ref{1004t}.
The probabilities $P_N(0)$, $P_N(N/2\sqrt{2})$, and $P_N(N/2)$ all fall
linearly with $1/N$, suggesting a constant weight if binned into intervals
whose width scales with $N$. It is worth noting that the weight at position
zero, which has the maximum number of interfering paths, does not
vanish completely due to destructive interference in any finite-length
QW. By contrast, the probability at positions $x = x_{\rm max}$ and $x =
N/\sqrt{2}$ scales as $P \propto N^{-2/3}$ [specifically, $P(x_{\rm max}) =
1.8 N^{-2/3}$ and $P(N/\sqrt{2}) = 0.44 P(x_{\rm max})$], accounting for the
sharpening of the distribution peaks with increasing $N$. We return to the
question of the peak shape in Sec.~IIIC. As noted in the preceding paragraph,
beyond $x = N/\sqrt{2}$ there is a very abrupt change in the form of $P(N)$
to an exponential decay, as shown in Fig.~\ref{1004t} for the point
$x = 0.7072N$ and in the sudden loss of oscillatory behavior in
Figs.~\ref{qob}(i) and \ref{qob}(j); this we will also analyze in more
detail in Sec.~IIIC.

\subsection{\label{sec2d3} Oscillatory Behavior}

\begin{figure}[t]
\includegraphics[width=0.48\textwidth]{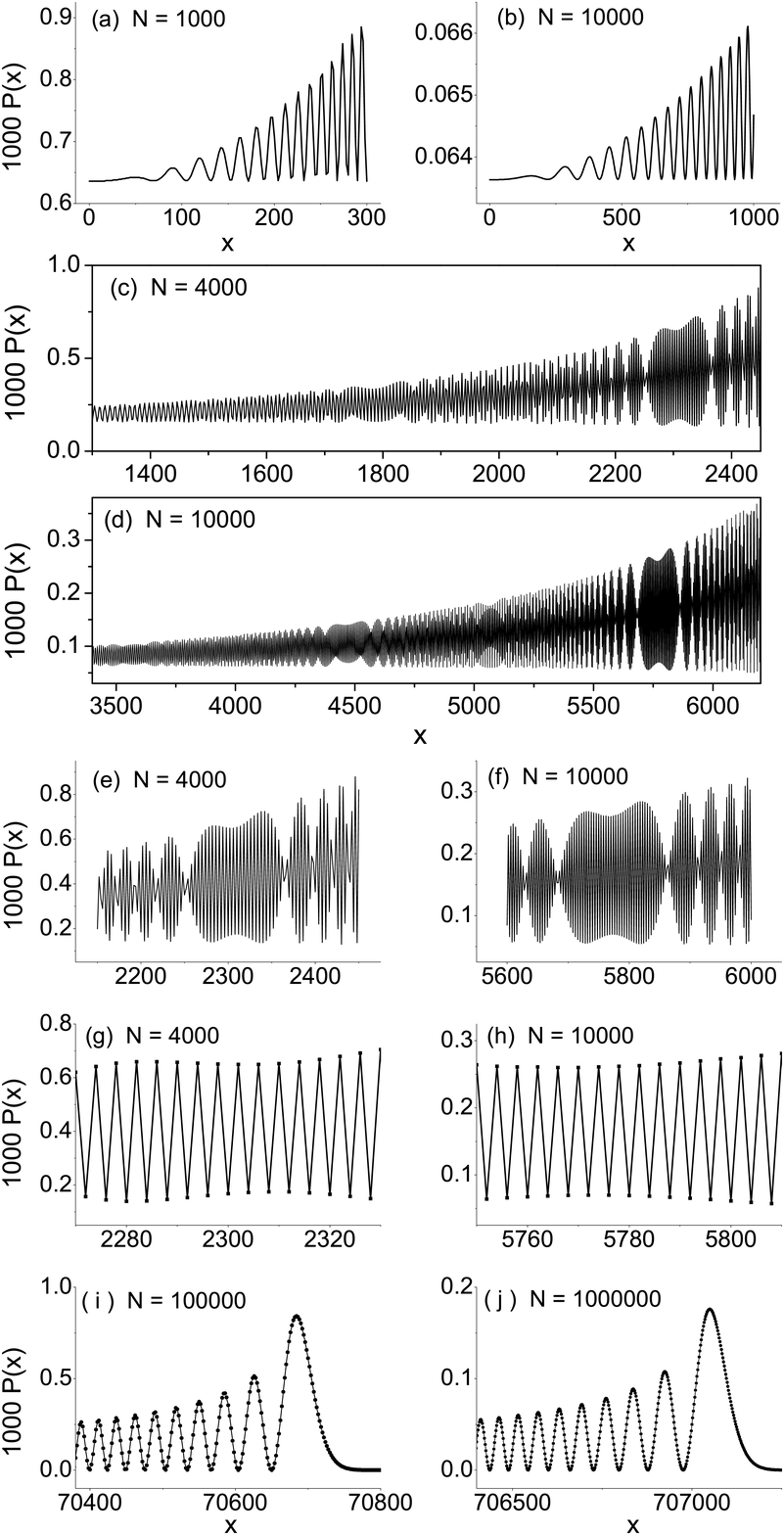}
\caption{Illustration of probability distributions in different parts of the
interval for 1D QWs of different step numbers $N$. First 15 oscillation
peaks for (a) $N = 1000$ and (b) $N = 10000$. Central region of the
distribution for (c) $N = 4000$ and (d) $N = 10000$, showing apparent
beating. Detail of the envelopes in the region of the beating structure
for (e) $N = 4000$ and (f) $N = 10000$. Detail within the envelope of the
beat structure for (g) $N = 4000$ and (h) $N = 10000$, showing the most
rapid oscillations in the distribution. Probability maxima for (i) $N =
100000$ and (j) $N = 1000000$. }
\label{qob}
\end{figure}

We turn next to the question of the oscillatory behavior of the probability
distribution within its envelope function. We stress again that this has
nothing to do with the period-2 oscillation created by the fact that walkers
alternate between odd and even sites at successive steps of the walk. We
begin by showing in Fig.~\ref{qob} the qualitative nature of the oscillations
in the real-space probability distribution function for selected regions of
the interval, using different values of $N$ to highlight their universal
nature.

\begin{table*}[bt]
\centering
\caption{Numbers of peaks in the probability distribution $P(x)$ found within
windows of width 100 steps for QWs of $N = 1000$, 10000, and 100000.}
\label{1005t}
\begin{tabular}{|c|c|c|c|c|c|c|c|}
\hline interval ($N$ = 1000) & [0,100] & [100,200] & [200,300] & [300,400] &
[400,500] & [500,600] & [600,700] \\
\hline number of peaks & 2 & 5 & 8 & 12 & 17 & 23 & 17 \\
\hline interval ($N$ = 10000) & [500,600] & [1500,1600] & [2500,2600] &
[3500,3600] & [4500,4600] & [5500,5600] & [6500,6600] \\
\hline number of peaks & 2 & 5 & 8 & 12 & 17 & 23 & 17 \\
\hline interval ($N$ = 100000) & [500,600] & [15500,15600] & [25500,25600] &
[35500,35600] & [45500,45600] & [55500,55600] & [65500,65600] \\
\hline number of peaks & 2 & 5 & 8 & 12 & 17 & 23 & 16 \\
\hline interval ($N$ = 10000) & [5000,5100] & [5500,5600] & [5600,5700] &
[5700,5800] & [5800,5900] & [5900,6000] & [6000,6100] \\
\hline number of peaks & 20 & 23 & 24 & 25 & 25 & 24 & 23 \\
\hline interval ($N$ = 100000) & [50000,50100] & [50000,50200] & [52000,52100]
& [54000,54100] & [56000,56100] & [58000,58100] & [60000,60100] \\
\hline number of peaks & 20 & 20 & 21 & 22 & 24 & 25 & 23 \\
\hline
\end{tabular}
\end{table*}

One of the most important properties of the oscillations is that their
effective wavelength, in terms of the fundamental step size, appears to
decrease towards larger values of $x$. At the center of the distribution,
as shown in Figs.~\ref{qob}(a) and \ref{qob}(b), they have relatively long
wavelengths, but towards the edges these become shorter [Figs.~\ref{qob}(c)
and \ref{qob}(d)] until in the region $x \approx 0.6N$ they have only twice
the fundamental length [Figs.~\ref{qob}(g) and \ref{qob}(h)]. Towards the
center of each half of the distribution, some beat-like structures develop
[Figs.~\ref{qob}(e) and \ref{qob}(f)]. As the peak is approached, the
frequencies drop and the oscillations vanish suddenly at $x = N/\sqrt{2}$
[Sec.~IIIA, Figs.~\ref{qob}(i) and \ref{qob}(j)]. We draw attention to the
fact that the distribution also has an effective lower envelope function,
in that the probability is never zero on any even points and in fact is
very much larger than the size of the oscillations close to $x = 0$
[Figs.~\ref{qob}(a) and \ref{qob}(b)], but we do not analyze this further
here.

\begin{figure}[b] \center
\includegraphics[width=0.48\textwidth]{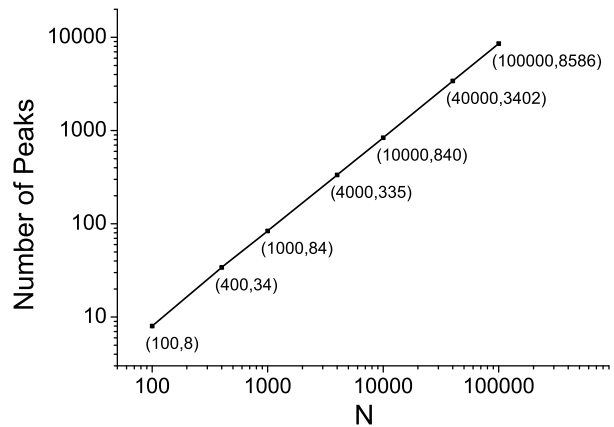}
\caption{Total number of peaks in the probability distribution for QWs
from $N = 100$ to 100000.}
\label{10051t}
\end{figure}

To quantify the nature of the oscillations, we count the numbers of peaks
in the probability distribution within different intervals and for QWs
of different $N$. First of all, by counting the total number of peaks it
is clear (Fig.~\ref{10051t}) that this is linearly proportional to
(approximately 8.5\% of) $N$. Subdividing the QW into intervals of fixed
length and counting the peaks in each of these gives the results shown in
Table \ref{1005t}. By comparing these peak counts horizontally, meaning
for different intervals within walks of the same $N$, a steady increase
in frequency becomes apparent out to $x \approx 0.6N$, from very
long-wavelength (periods of 50 or more steps) oscillations near $x = 0$
to extremely rapid (period-4) ones when $x$ is a signifcant fraction of
$1/\sqrt{2}$. We remind the reader here that a period of 4 in a system
where odd sites have probability zero is essentially a max-min-max-min-$\dots$
structure within the envelope [Figs.~\ref{qob}(g) and \ref{qob}(h)]. Thus
spatial information about the QW is truly contained on all length scales.
By comparing the peak counts vertically, meaning for different values of
$N$, it becomes apparent that corresponding regions have precisely the same
frequencies, with the maximum frequency occurring in the region around $x =
0.58N$. Thus the spatial modulation of the QW is a quantity independent of
$N$; although QWs of different $N$ cannot be called self-similar, they do
share similarities in particular aspects.

Other forms of similarity and scaling appear in particular segments of the
probability distribution. Considering first the region close to $x = 0$,
Fig.~\ref{qob}(a) shows 15 peaks in the region $[0,300]$ for $N = 1000$.
The corresponding region for $N = 10000$, which is $[0,3000]$, contains
148 peaks, confirming the conclusion drawn above that the frequencies
are the same in corresponding areas for differing $N$, leading to the
overall linearity in $N$ of the peak number (Fig.~\ref{10051t}). However,
by counting the first 15 peaks in the probability distribution for $N =
10000$, shown in Fig.~\ref{qob}(b), they fill the interval $[0,960]$,
indicating a $\sqrt{N}$ scaling of the maximum wavelengths around the
center of the distribution. We clarify that this is not in contradiction
with Table \ref{1005t}, where the representative low-$x$ intervals are
taken at different finite values of $x$.

This type of scaling may also be observed in other regions of the QW
probability distribution. Focusing next on the special structures noted
above, Figs.~\ref{qob}(c) and Figs.~\ref{qob}(d) show three of these, which
we find quite reproducibly in the region around $x = N/2$. More accurately,
these macroscopic dips of the distribution envelope appear around  $x =
0.36N$, $x = 0.45N$, and $x = 0.58N$. The interval around $0.58N$, which is
also the region with maximum oscillation frequency, shows a particularly
remarkable beating structure [Figs.~\ref{qob}(e) and \ref{qob}(f)], with
multiple points where the upper and lower envelopes meet. Figures \ref{qob}(g)
and \ref{qob}(h) show the maximal frequency regime and parts of the beating
envelopes in the fullest detail, and Table \ref{2001t} contains the
corresponding information for the beating interval for QWs of $N = 1000$,
10000, and 100000. Again it is clear that a factor-10 increase in $N$ causes
only a factor $\sqrt{10}$ magnification of the width of the beating structure
and, given the fixed maximal frequency in this interval, of the number of
peaks it contains.

\begin{table}[b]
\centering
\caption{Position, width, and number of peaks in the distinctive beating
structure around $x = 0.58N$ for walks of $N = 1000$, 10000, and 100000 steps.}
\begin{tabular}{|c|c|c|c|}
\hline N & region & width & number of peaks \\
\hline 1000  & [546,604]    & 58 & 14 \\
\hline 10000 & [5682,5860]  & 176 & 44 \\
\hline 100000 & [57452,58016] & 566 & 141 \\
\hline
\end{tabular}\label{2001t}
\end{table}

\begin{figure}[t] \center
\includegraphics[width=0.45\textwidth]{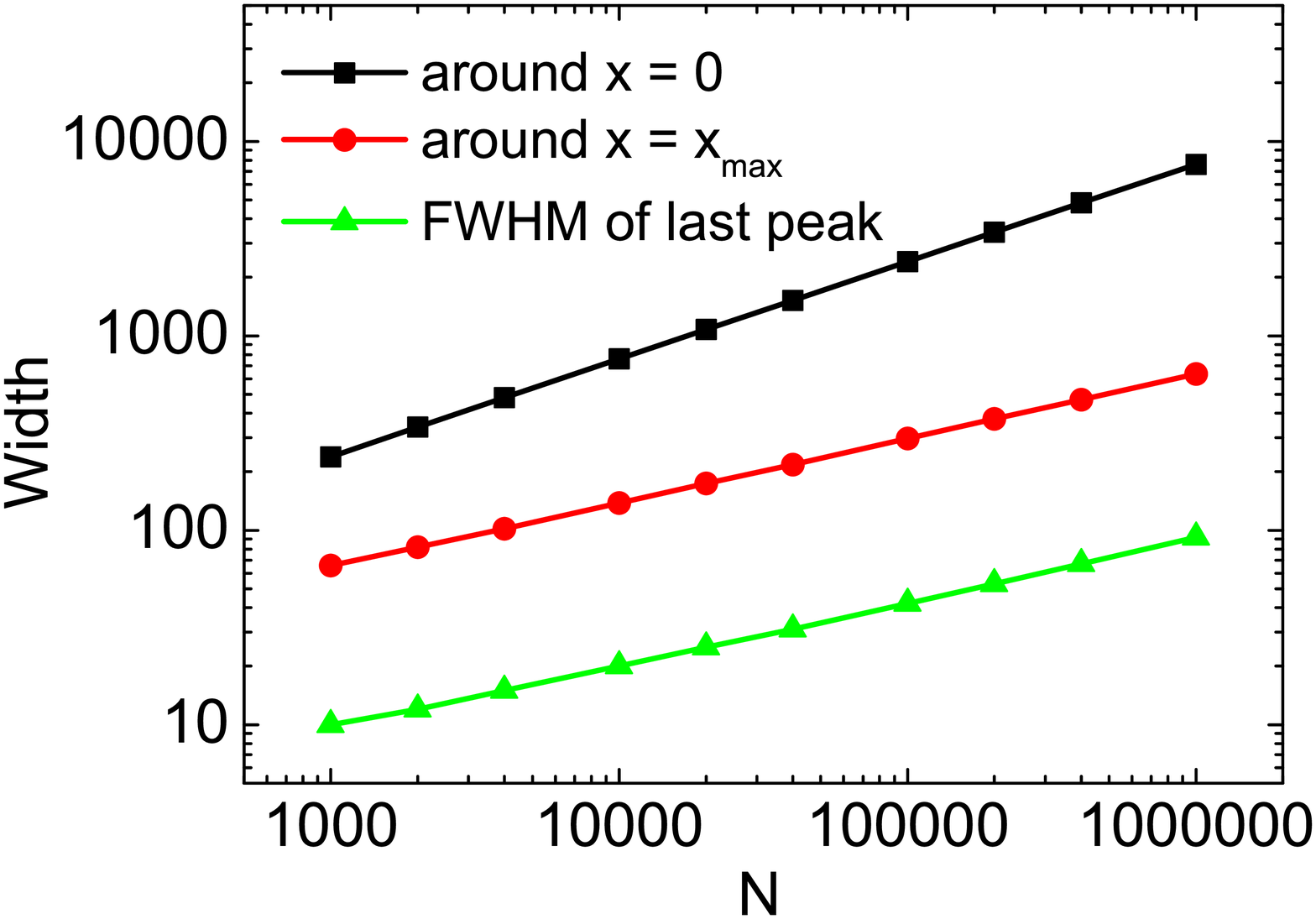}
\caption{Width in $x$ of different numbers of peaks in the probability
distribution for QWs over the full range of $N$ values studied. The width
of the first 10 peaks from $x = 0$ (black) scales with $\sqrt{N}$; the width
of the last 10 peaks up to the maximum (red) and also the FWHM of the tallest
peak (green) both scale with $N^{1/3}$.}
\label{qop}
\end{figure}

We close our discussion of the scaling of peak widths by considering the
probability oscillations around $x = N/\sqrt{2}$. By counting the width of
the region covered by the last 10 peaks up to and including the peak of
maximum probability, we find that this scales according to $N^{1/3}$
(Fig.~\ref{qop}). Similarly, the full width at half maximum height (FWHM)
of the leading peak in $P(x)$ also scales with $N^{1/3}$, and hence this
last peak retains its aspect ratio when $x$ is normalized by $N$. Thus we
demonstrate the presence of algebraic scaling in the probability oscillations
over the full distribution.

\subsection{\label{sec2d6} Peak Shape}

Next we consider the shape and scaling of the asymmetric envelope of peaks in
the probability distribution to ascertain its functional form, $P(x,N)$. The
dependence on $N$ is largely contained in Fig.~\ref{1004t} and we make these
results more systematic here. Concerning the dependence on $x$, we take the
results of Sec.~IIIA as a demonstration that the ``crossover'' region just
beyond the maximum peak becomes a set of vanishing measure at large $N$; to
fit the envelope function, meaning the set of points extracted from the full
data set that fall close to the upper edge of $P(x)$, for the region $x \le
x_{\rm max}$, we assume that it diverges at $x = \pm \, N/\sqrt{2}$ for large
values of $N$. Before considering the QW, we recall that the functional form
of the probability distribution for a classical random walk (Sec.~IIA) becomes
a Gaussian at large $N$, with the form
\begin{equation}
P(x) = P_0 + A e^{-\frac{(x - b)^2}{2 \sigma^2}},
\label{egf}
\end{equation}
where $\sigma = \sqrt{N}$, $A = 1/\sqrt{2 \pi N}$, and $P_0 = 0 = b$ for a
normalized and centered distribution. This is an exponential function whose
characteristic width scales with $\sqrt{N}$.

\begin{figure}[t]
\includegraphics[width=0.45\textwidth]{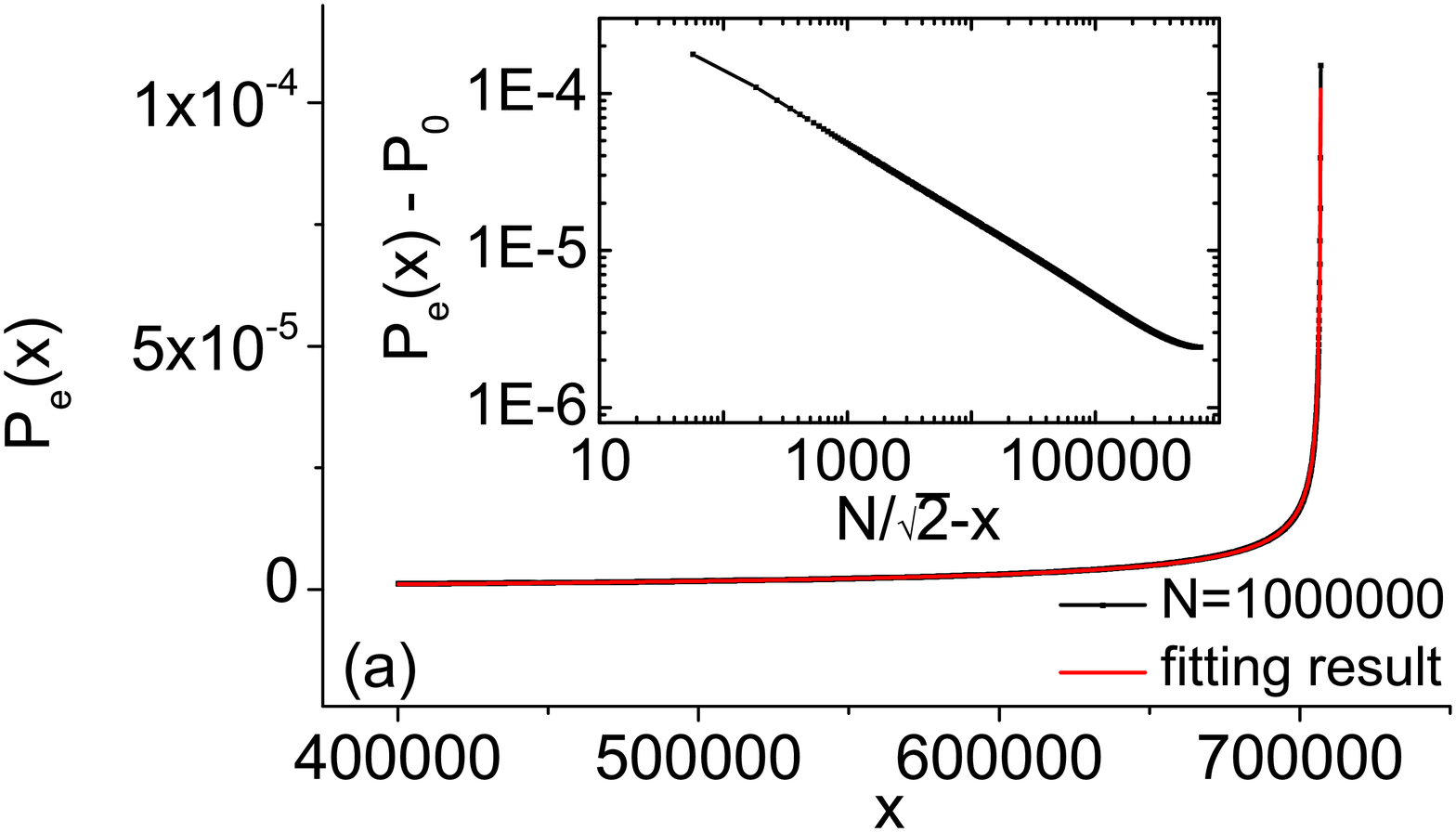}
\includegraphics[width=0.45\textwidth]{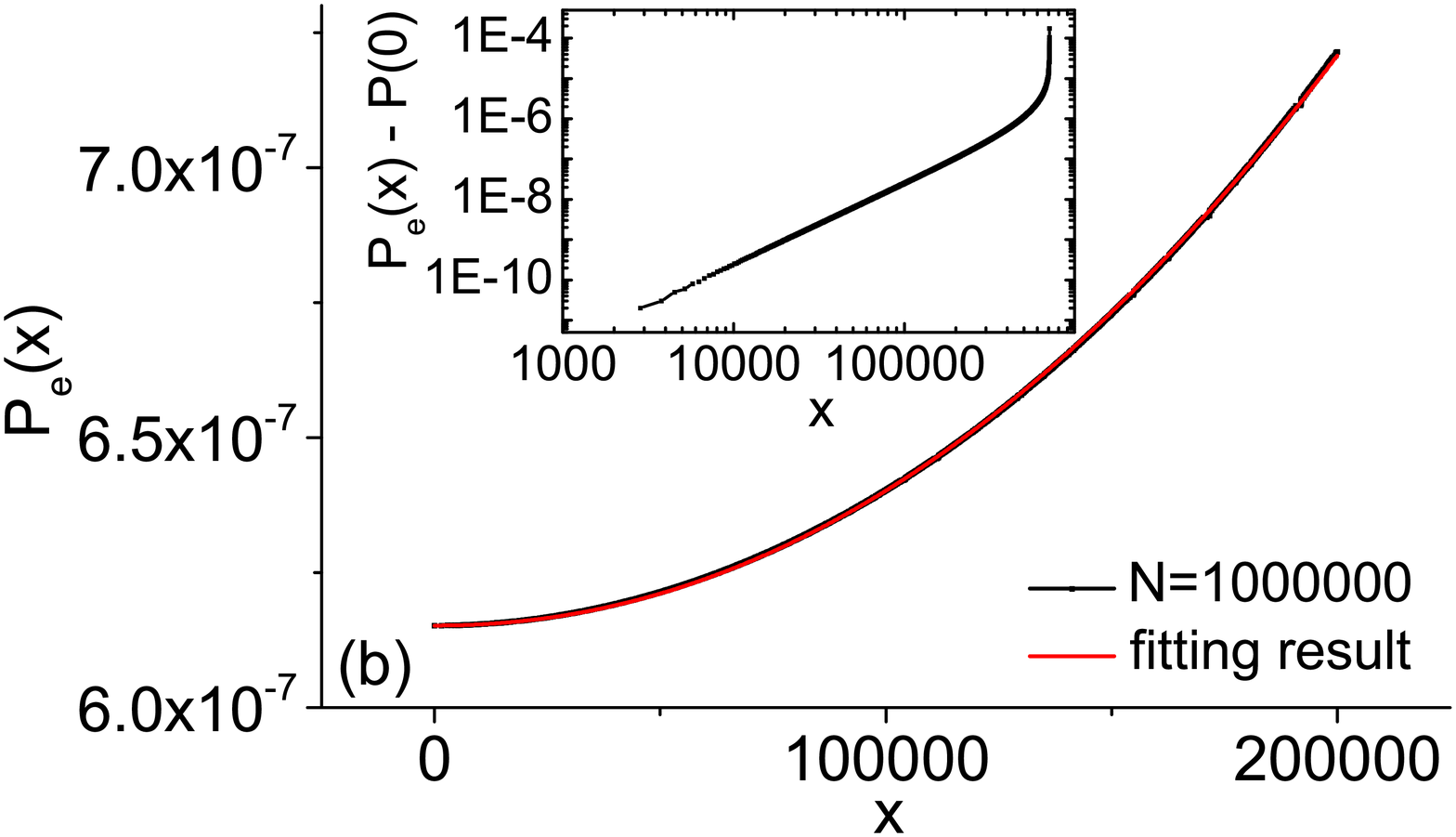}
\includegraphics[width=0.45\textwidth]{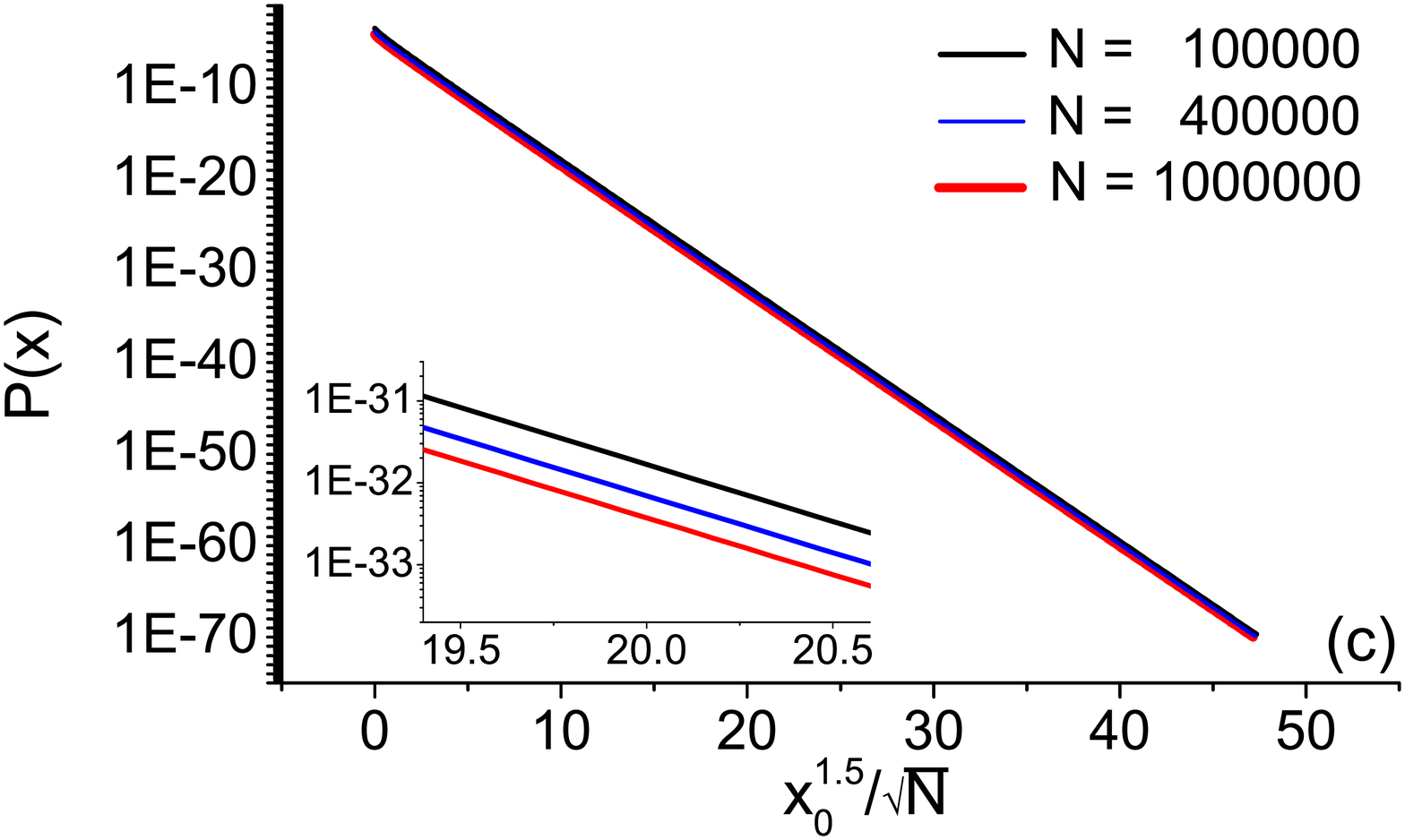}
\caption{Fit to the envelope of the probability distribution, $P_e (x)$, for
a QW of $N = 1000000$ steps. (a) Fit for the interval $400000 \le x \le
x_{\rm max}$ using the algebraic form of Eq.~(\ref{01e}) with $B = 1.884$,
$A = 1.487$, $b = 707144$, and $c = 0.4918$ (see text). (b) Fit for the
interval $0 \le x \le 200000$ using the algebraic form of Eq.~(\ref{02e})
with $B' = 0.6152$, $A' = 1.0017$, and $c' = 2.074$ (see text). (c) Fit to
the probability distribution $P(x)$ for the interval $x > N/\sqrt{2}$ using
the exponential form of Eq.~(\ref{03e}) with $P_0 = 0$, $b = N/\sqrt{2}$,
$d = 3.2$, and $A = 0.8$ (see text).}
\label{fit01}
\end{figure}

By contrast, for the QW with large $N$ we find an excellent fit to an
algebraic form,
\begin{equation}
P_e (x) = P_0 + a (b - x)^{-c}
\label{01e}
\end{equation}
for each half of the distribution, with $b = \pm \, N/\sqrt{2}$. To quantify
the extent of the validity of such a fit, we examine the data on logarithmic
axes [inset, Fig.~\ref{fit01}(a)], finding that a single power-law provides
a robust description of the envelope for the entire region $0.4N \le x \le
x_{\rm max}$. For the three free parameters, our results as $N \rightarrow
\infty$ (in practice, up to $N = 1000000$) indicate that the constant $P_0 =
- B/N$ with $B = 1.884$. Fits performed using only two remaining free
parameters, and illustrated in the main panel of Fig.~\ref{fit01}(a) for
the case $N = 1000000$, allow us to deduce for the large-$N$ limit that
the prefactor approaches $a = A/\sqrt{N}$ with $A = 1.5$ and the exponent
approaches $c = 0.5$. We conclude that to an excellent approximation the
envelope function follows a square-root form in $x$ measured away from
$x = \pm \, N/\sqrt{2}$, and this determines the algebraic form of the
peak width. The behavior of the prefactor $A$ ensures that $P_e (x)
\propto 1/N$ across all of this range, consistent with the results of
Fig.~\ref{1004t} but excluding the final peak. We remind the reader that
the envelope function $P_e(x)$ is not a quantity obeying a normalization
law as the true distribution $P(x)$ does.

Below $x = 0.4N$, the shape of the envelope begins to deviate from the
universal square-root form [inset, Fig.~\ref{fit01}(a)]. To ascertain its
shape close to the center of the distribution, we instead apply a fit of the
form
\begin{equation}
P_e (x) = P_0 + a' x^{c'}
\label{02e}
\end{equation}
for each half of the distribution. Again the data on logarithmic axes [inset,
Fig.~\ref{fit01}(b)] show an excellent fit to a single set of parameters over
a broad region, $0 \le x \le 0.2N$, where $P_0 = B'/N$ with $B' = 0.615$,
$a' = A'/N^{c'+1}$ with $A' = 1$, and $c' = 2$. The results are shown in
the main panel of Fig.~\ref{fit01}(b). Thus the $P \propto 1/N$ form is
maintained, the universal behavior of the envelope around its center is a
quadratic dependence on $x$, and there is a relatively significant constant
contribution that reflects directly the incomplete nature of destructive
interference in the central region of the QW. We regard the intermediate
regime $0.2N \le x \le 0.4N$ as a crossover zone between the two limiting
forms [Eqs.~(\ref{01e}) and (\ref{02e})] and do not consider it further.

We close our discussion of the probability distributon $P(x,N)$ by considering
the region $x > N/\sqrt{2}$. Here there are no longer any oscillations
[Figs.~\ref{qob}(i) and \ref{qob}(j)] and $P(x,N)$ is the ``envelope.''
In Sec.~IIIA [Fig.~\ref{1004t}] we showed that the dependence on $N$
crosses very rapidly to an exponential decay around $x = N/\sqrt{2}$. A
complete fit of the data in this regime reveals the form
\begin{equation}
P(x) = P_0 + a e^{- d \frac{(x - b)^{1.5}}{N^{0.5}}},
\label{03e}
\end{equation}
with $P_0 = 0$, $b = N/\sqrt{2}$, $d = 3.2$, and $a = A N^{-2/3}$ with $A =
0.8$, i.e.~an exponential and pseudo-Gaussian behavior but with unconventional
alterations to the exponents in both $x$ and $N$. The effectiveness of this
fit is shown in Fig.~\ref{fit01}(c), which also highlights how rapidly the
probability falls away in a short distance beyond $x = N/\sqrt{2}$.

\subsection{\label{sec2r} Fourier Transformation of the 1D-QW }

In Sec.~IIIB we explored the rich spatial frequency information contained
in the oscillations of the probability distribution $P(x)$. The QW contains
oscillations on all length scales from ultra-short wavelengths around $x =
0.58 N$ to long-wavelength oscillations scaling as $\sqrt{N}$ around $x = 0$,
with similar algebraic scaling around $x = x_{\rm max}$. These differing spatial
frequencies can even combine to create highly reproducible beating structures
in certain regions. All of this information should be reflected in the Fourier
transform of $P(x)$, which we discussed from an analytical point of view in
Sec.~IIC.

Here we perform a discrete Fourier transformation on the data sets for 1D
QWs of all lengths $N$, finding results for the Fourier components, $F(k)$,
of the type shown already in Fig.~\ref{num-ana}(b) for $N = 100$ and $N =
1000$. $F(k)$ possesses a primary peak with amplitude $F(0) = 1$ at $k = 0$,
flanked by two secondary peaks with negative components at $k \simeq \frac{5
\pi}{N}$, and then shows an oscillatory form between positive and negative
values of the Fourier components. The oscillations are again contained within
a decaying envelope function, which we find to be identical at positive and
negative values, and the $k$-space periodicity of the oscillation is remarkably
constant across the range $- \pi < k \le \pi$. This result is a clear
reflection of the fact that spatial information is present in the probability
distribution on all length scales, and the mixture of positive and negative
components across the range of $k$ is manifest in complex mixing phenomena
such as the beating structure. However, beyond the large $k = 0$ component
there are no special spatial frequencies appearing in the distribution. We
comment here that the constant component $F(0) = 1$ is simply the sum of all
data in real space, and therefore is the result expected for a normalized
distribution.

\begin{figure}[t]
\includegraphics[width=0.5\textwidth]{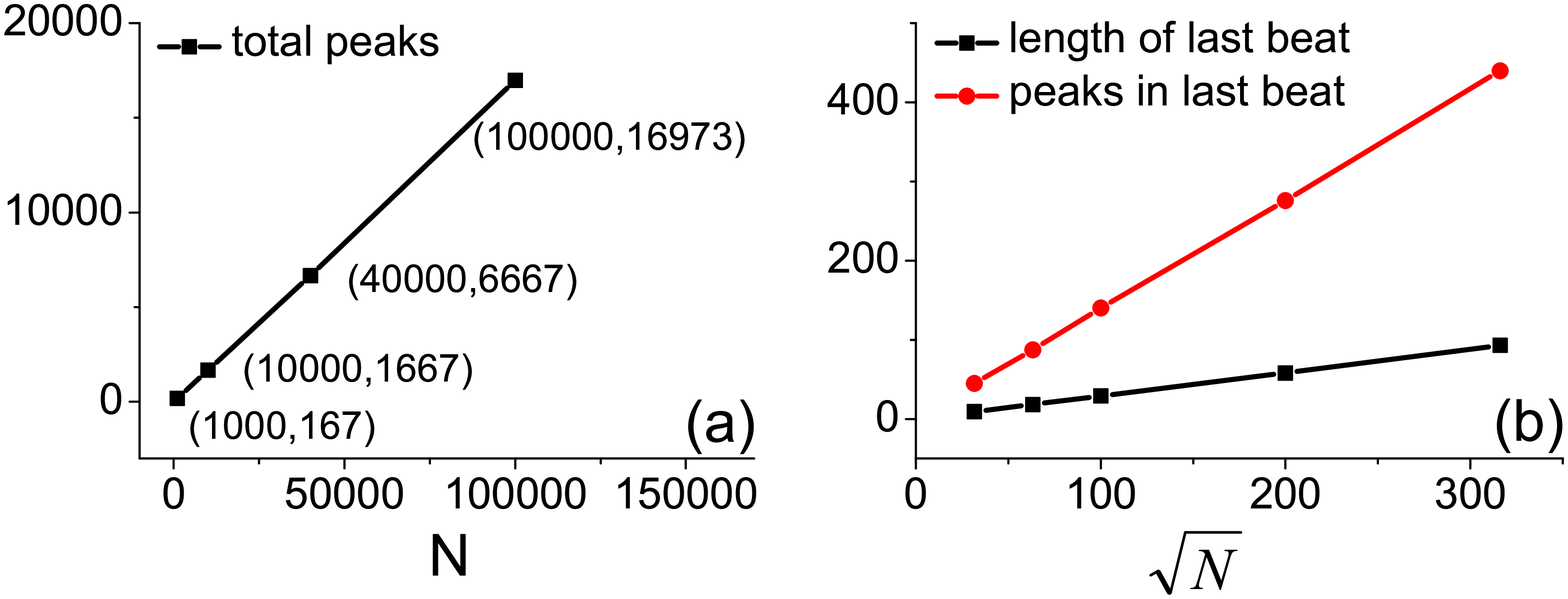}
\caption{(color online) (a) Number of peaks in Fourier space for QWs of
lengths $N = 100$ to 100000. (b) Length of the envelope function of the
final beat below $k = \pi$, in units of $Nk/2\pi$ (black), and the number
of peaks in this region (red), shown as a function of $\sqrt{N}$.}
\label{fourier01c}
\end{figure}

\begin{figure}[t]
\includegraphics[width=0.48\textwidth]{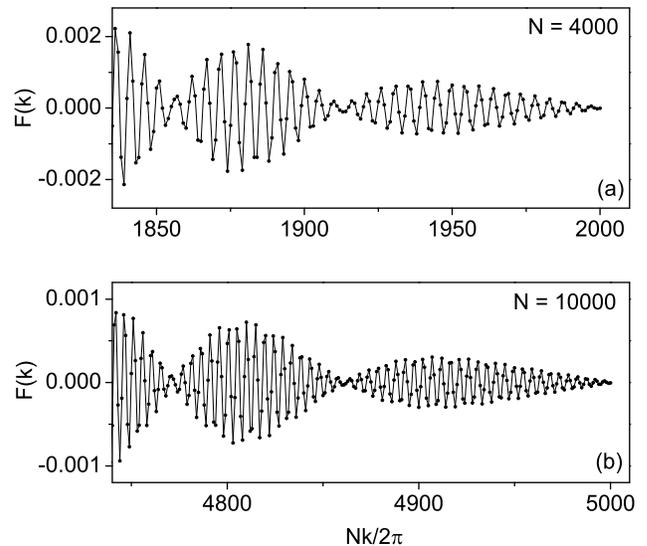}
\caption{Fourier components of the probability distribution close the maximum
frequency, shown rescaled to $Nk/2\pi$. In this regime the envelope function
shows clear beating behavior.}
\label{fourier01a}
\end{figure}

We begin our quantitative analysis of the Fourier transform $F(k)$ by
counting its peaks. Figure \ref{fourier01c}(a) confirms that the total
number of peaks in the Fourier spectrum scales linearly with $N$, as in
real space and again with a constant of proportionality of order 8\%
(more precisely, 1/12) for each half of the transformed distribution.
As in real space, we may also count the peaks in particular parts of
the distribution to investigate their scaling form. The most striking
feature of the Fourier transformed data for large values of $N$ is a
reappearance of beating phenomena between the upper and lower envelope
functions. In complete consistency with the results of Sec.~IIIB, where
the beating structures were observed in the region of the distribution
with the highest spatial frequencies, the Fourier-space beats are
clearest close to $k = \pi$. In Fig.~\ref{fourier01a} we illustrate this
property with the $k$-axis rescaled to $Nk/2\pi$ to better reflect the
number of Fourier components in the data set. As in Sec.~IIIB, we may
characterize the structure of the beat pattern by considering the length
of the final beat and the number of peaks it contains, which are tabulated
in Table \ref{tfour} and illustrated in Fig.~\ref{fourier01c}(b). From the
latter it is clear once again that the beat structures scale according to
$\sqrt{N}$.

\begin{table}[b]
\centering
\caption{Characterization of oscillation frequencies in Fourier space. The
length of the last beat is quoted in units where the interval of the Fourier
components is rescaled to $(- Nk/2\pi,Nk/2\pi]$.}
\begin{tabular}{|c|c|c|c|}
\hline N      & total peaks & peaks in last beat & length of last beat \\
\hline 1000   & 167         & 9                & 45 \\
\hline 4000   & 667         & 18               & 87 \\
\hline 10000  & 1667        & 29               & 140 \\
\hline 40000  & 6667        & 58               & 276 \\
\hline
\end{tabular}
\label{tfour}
\end{table}

\begin{figure}[t]
\includegraphics[width=0.45\textwidth]{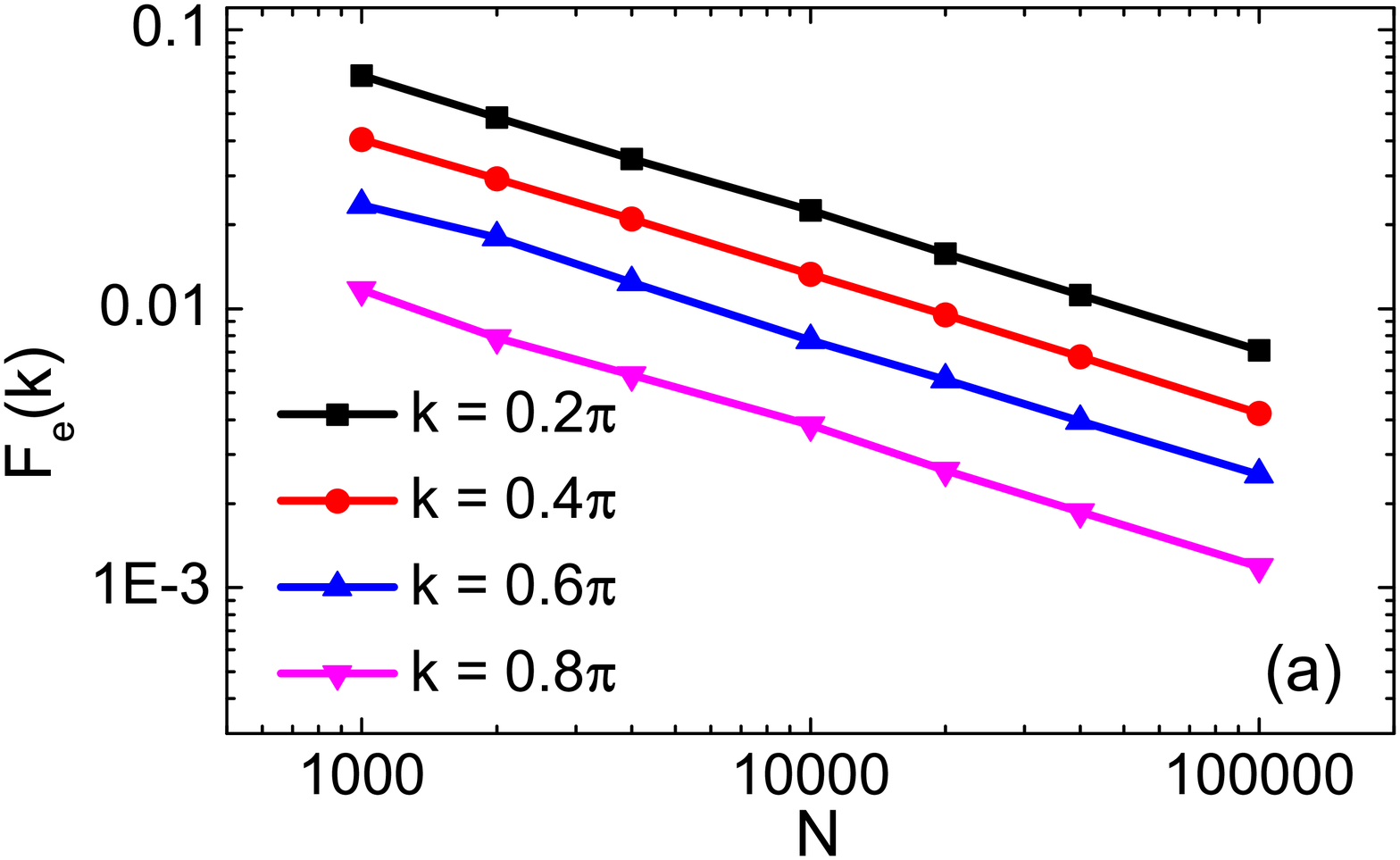}
\includegraphics[width=0.45\textwidth]{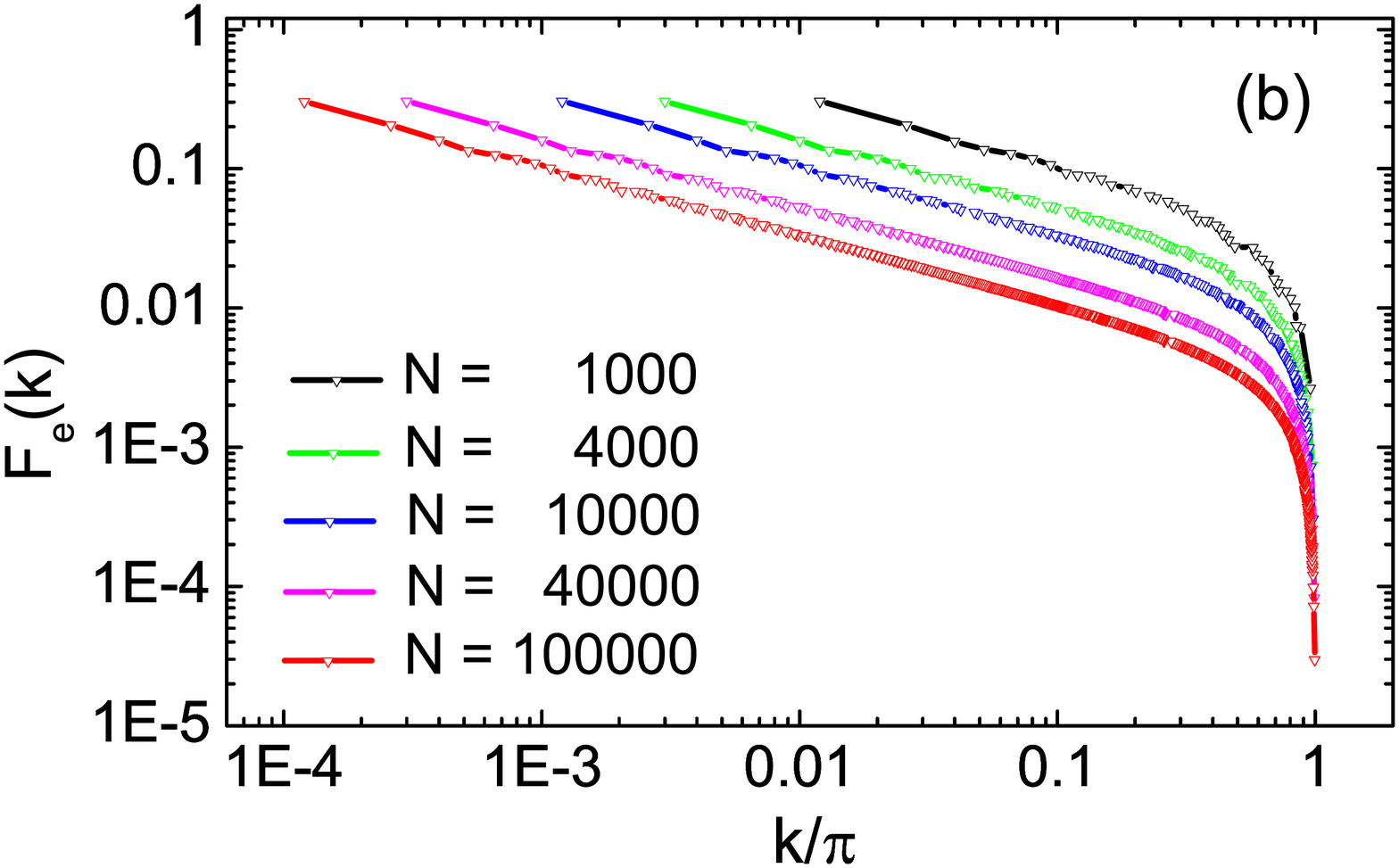}
\includegraphics[width=0.45\textwidth]{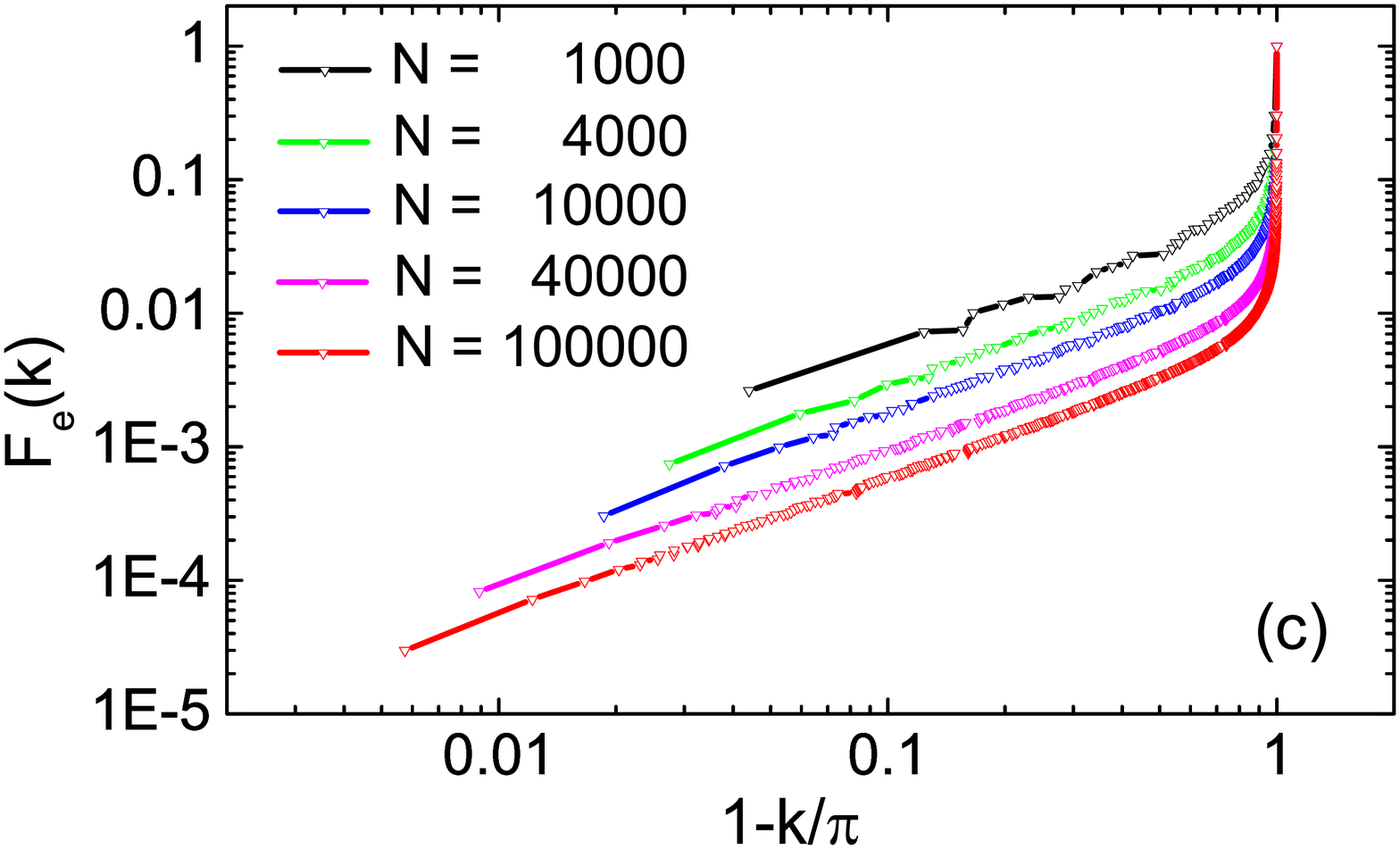}
\caption{Fit to the envelope of the Fourier transform of the probability
distribution, $F_e (k)$, for QWs of $N = 1000$ to 100000 steps. (a)
$N$-dependence of $F_e (k)$ for four different values of $k$, showing a
clear algebraic decay with exponent 1/2. (b) Fit using the algebraic form
of Eq.~(\ref{04e}), demonstrating its relevance for the interval $0 \le k
\le 0.2 \pi$ and returning the exponent $c = 1/2$. (c) Fit using the
algebraic form of Eq.~(\ref{05e}), demonstrating its relevance for the
interval $0.4 \pi \le k \le \pi$ and returning the exponent $c' = 1$. }
\label{fit02}
\end{figure}

Proceeding as in Sec.~IIIC, we consider the possibility of a universal
fit to the envelope function $F_e(k,N)$, whose $k$-dependence is valid
for all values of $N$. The $N$-dependence of $F_e (k,N)$ is shown in
Fig.~\ref{fit02}(a) for values of $k$ from across the full range, and
the constant slope gives the clear result $F_e (k,N) \propto 1/\sqrt{N}$.
Mindful of the fact that the real-space envelope changes form between the
limits of small and large $x$, we consider the functional forms
\begin{equation}
F(k) = \frac{A}{\sqrt{N}} \left( \frac{k}{\pi} \right)^{-c}
\label{04e}
\end{equation}
and
\begin{equation}
F(k) = \frac{A'}{\sqrt{N}} \left( 1 - \frac{k}{\pi} \right)^{c'}
\label{05e}
\end{equation}
for small and large $k$. As shown in Figs.~\ref{fit02}(b) and \ref{fit02}(c),
Eq.~(\ref{04e}) with exponent $c = 1/2$ provides an excellent fit at small
$k$, out to $k \approx 0.2 \pi$, and Eq.~(\ref{05e}) with $c'= 1$ an excellent
fit for all $k$ values in the upper half of the range. Thus we find the
$k$-dependence of the envelope of Fourier components to be algebraic over
the whole range, with an inverse square-root decay away from $k = 0$ crossing
over to a linear decrease as $k$ approaches $\pi$.

In summary, the Fourier transform of the QW probability distribution contains
all of the same information in a complementary form. It is bounded by upper
and lower envelope functions with the same algebraic decay. It demonstrates
that spatial frequencies are present on all scales from the inverse step
length to the inverse system size, with no special internal period(s) but
with distinctive beating structures on a length scale of $\sqrt{N}$. While
the QW does not satisfy the strict definition of self-similarity (no fractal
structures appear), many of its features are similar and scale-invariant
across the full range of $N$ values. Thus the simple quantum evolution
algorithm of Sec.~IIB contains a very rich variety of spatial information.

\section{\label{sec4} Two-Dimensional Quantum Walks}

\subsection{\label{sec4a} Classical Random Walk in 2D}

As noted in Sec.~I, it is the advent of experiments capable of realizing
a 2D QW \cite{Schreiber06042012} that has caused the resurgence of interest,
both experimental and theoretical, in the field. Before discussing the range of properties exhibited by QWs in 2D,
it is helpful to review the classical case. In principle there are
two ways to generalize the unbiased 1D classical random walk to 2D.

\noindent
1) Adopting a square grid, the walker is equipped only with a
two-face coin and therefore flips it once to step to $(x \pm 1,y)$, then
a second time to arrive at $(x \pm 1,y \pm 1)$. Now the conventional $N$-step 1D binomial
distribution for $x$ or $y$ is recovered by summing over the probabilities
in the orthogonal direction and it is easy to show for any $N$ that
$P(x,y) = P(x)P(y)$. In the large-$N$ limit, the distribution approaches
\begin{equation}
P(x,y) = {\textstyle \frac{1}{2 \pi N}}e^{-\frac{x^2 + y^2}{2N}}
\label{ep2dcrw2}
\end{equation}
and the mean distance of the walker from the origin after N steps is $\langle r \rangle = \sqrt{N}$.

\noindent
2) Remaining on a square grid, the walker at site $(x,y)$ has equal
probabilities of 1/4 (equivalent to a four-face coin) to move to any of
the points $(x \pm 1, y \pm 1)$. It results in an identical distribution to case (1).

A further generalization is to consider a continuous space in which
the walker takes unit-length steps at any random angle $0 \le \phi < 2\pi$.
The walk approaches perfect circular symmetry ($r^2 = x^2 + y^2$) at large values of $N$.

\subsection{\label{sec4b} Two Types of 2D QW}

With a view to experimental realization, 2D QW also has two-face coin scheme
and four-face coin scheme.

\begin{figure}[htpb]
\centering
\includegraphics[width=0.5\textwidth]{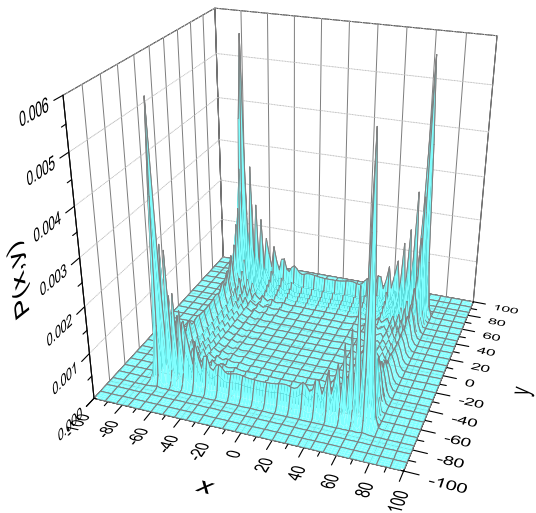}
\caption{Probability distribution of 2D QW effected using a four-face
coin selecting steps in the $\pm$($x$+$y$) and $\pm$($x$-$y$) directions,
for $N = 100$ steps.}
\label{fig:four}
\end{figure}

A four-face coin $H_4$ can be constructed by
product the Hadamard operators, i.e. $H_4=U_H\otimes U_H$(as well as corresponding initial state).
It evolves to a distribution shown in Fig.~\ref{fig:four} for a walk of 100 steps.
We note immediately that this 2D QW is a directly expand of 1D QW, in which $P(x,y)=P(x)P(y)$.
One will understand this equivalence by invoking the matrix identity
\begin{equation}
(A_1B_1)\otimes(A_2B_2)=(A_1\otimes A_2)(B_1\otimes B_2),
\label{otimes1}
\end{equation}
$A_1, A_2$ represent the Hadamard operator $U_H$ and $B_1, B_2$ for the initial state
 $|\uparrow\rangle + i|\downarrow\rangle$.

\begin{figure}[htpb]
\centering
\includegraphics[width=0.5\textwidth]{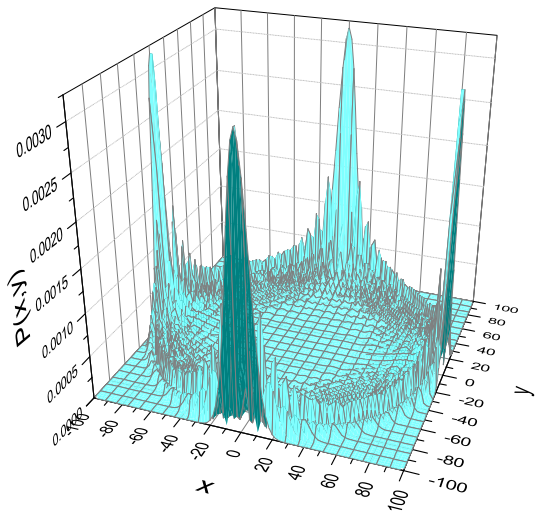}
\caption{Probability distribution of 2D QW effected using a two-face coin
twice in each cycle to select steps in the $\pm x$ and $\pm y$ directions,
for $N = 100$ cycles.}
\label{fig:2dqrw2cc}
\end{figure}

The two-face-cion scheme lead to a completely different 2D distribution, shown in
Fig.~\ref{fig:2dqrw2cc} for a walk of 100 cycles. Specifically, an evolution
protocol using the same two-face coin twice in each complete cycle to
generate successive steps in the $x$ and $y$ directions. The first flip
of the coin selects the direction of $\pm x$ and the second $\pm y$.
A full cycle can be represented as $\hat U = S_y Y (S_x Y)$ and the quantum
state after $N$ evolution cycles as $|\psi_N \rangle = (\hat U)^N |\psi_0
\rangle$.This probability distribution has several properties in common with the 1D QW,
including strong probability peaks far from the center, strong destructive
interference everywhere near $(x,y) = (0,0)$, and oscillatory behavior
around the locus of maxima. However, the qualitative features of this QW
are strikingly different from those of the four-face-coin walk, most notably
in that the peaks are rotated by 45$^{\rm o}$, they occur right at the edge of
the system, and here the locus of maxima does show some tendencies towards a
circular shape, even if the peaks upon it remain strongly anisotropic and
four-fold symmetric. The location of the probability maxima is really very
anomalous, in that the minimum of the destructive interference is truly
obtained when a walker makes all $N$ of its steps in the same direction
for one lattice orientation, but precisely $N/2$ in each direction for
the other orientation.


Fig.~\ref{fig:2dqrw2cc} can also be realized by a four-face coin(Grove walk\cite{a4}).
In one of the most notable recent studies \cite{a8,a9}, Di Franco and
coauthors proved the equivalence of the spatial probability distributions
between the 2D QW using a single two-face coin, which these authors termed the
``alternate quantum walk'' (AQW) and the Grover walk. The AQW is actually a 2D QW
 in which the two direction are maximal entangled.

The Grover coin can be expressed in the form
\begin{equation}
G_4 = \frac{1}{2} \left( \! \begin{array}{cccc} -1 & 1 & 1 & 1 \\
1 & -1 & 1 & 1 \\ 1 & 1 & -1 & 1 \\ 1 & 1 & 1 & -1 \end{array} \! \right) \! ,
\label{004}
\end{equation}
Noted that the distribution and spreading rates are effected by the initial state \cite{a3,a4}.
Eq.\ref{e104} is the one providing Grove walk a maximum spread rate
\begin{equation}
|\psi_0^m \rangle = {\textstyle \frac{1}{2}} (|\uparrow \rangle -
|\rightarrow \rangle - |\leftarrow \rangle + |\downarrow \rangle)
|0_x,0_y \rangle.
\label{e104}
\end{equation}

We conclude here the difference between the two type of 2D distribution,
is not distinguished by two-face coin or four-face coin, but whether the two orthogonal directions are entangled.
Fig.~\ref{fig:four} can also be realized using two uncorrelated two-face coins.
Some authors have studied decoherence \cite{a5} and localization \cite{a7}
in 2D QWs using the concept of two coins, and others have quantified the
effects of entanglement by a partial or complete swapping of the two coins
after each step of the walk \cite{a6}.

The method of generating a QW by using the same two-face coin twice, to
which we refer henceforth as the AQW \cite{a8,a9}, is an important one for
a number of reasons. First and foremost is that two-state systems are much
easier to find, or to produce, in any physical realization of a quantum
walker, and hence are much more relevant for experimental implementation.
Secondly, this maximally entangled protocol contains further unconventional
phenomena, which we discuss in Sec.~V. Further, coin entanglement provides
a valuable and completely general means of constructing and perhaps also of
realizing high-dimensional QWs using only two-face coins.

We conclude this subsection by contrasting the two QWs generated by
unentangled and by maximally entangled coins. The unentangled
system appears to show a perfectly square, factorized probability distribution
(a result we demonstrate in Sec.~V) characteristic of the 1D QW, with
maxima at positions converging to $(\pm \, N/\sqrt{2}, \pm \, N/\sqrt{2})$.
The entangled case shows a circular ``probability front'' and significant
complexity in the interference pattern within it. In the sense that the
unentangled situation can be discussed as accelerated diffusion from the
classical random walk to the quantum walk as a result of destructive interference,
so the maximally entangled walk appears to show a still further accelerated
diffusion. Indeed, it achieves a situation where information propagates to
the very edge of the system with high probability, which is a quite
remarkable consequence of near-perfect destructive interference among
trajectories in the center region. A possible interpretation of this
result may be found in preservation of the information content of the
coin, because the degree of two-coin entanglement is ``complete'' in
the sense that the walk can be generated using only one coin.

\section{\label{sec5} Numerical Studies of 2D QWs}

For an exact characterization of the 2D QWs introduced in Sec.~IVB, we
turn now to a numerical investigation of their probability distributions.
We begin by contrasting in Table \ref{label101} the analytical results
for the probabilities of the two walks for $N = 6$ steps.

\begin{table}[b]
\caption{Probability distributions of the 2D QW for $N = 6$ steps, shown
as $4096 P(x,y)$. On the left is the four-face-coin QW (equivalent to two
unentangled coins) and on the right the AQW (equivalent to a single coin,
or two maximally entangled coins). For the interpretation of these numbers
we note that in the 1D QW with $N = 6$ one obtains $64 P(x) = [1, 18,
9, 8, 9, 18, 1]$.}
\begin{equation*}
\begin{tabular}{|c|ccccccc||ccccccc|}
\hline \backslashbox{y\!\!}{\!\!x} & -6\! & -4\! & -2\! & 0\! & 2\! & 4\!
& 6 & -6\! & -4\! & -2\! & 0\! & 2\! & 4\! & 6 \\
\hline
6 & 1 \!& 18\!& 9 \!& 8 \!& 9 \!& 18\!& 1 & 1 \!& 26\!&125\!&200\!&125\!&
26\!& 1  \\
4 & 18\!&324\!&162\!&144\!&162\!&324\!& 18& 26\!& 68\!& 50\!&208\!& 50\!&
68\!& 26 \\
2 & 9 \!&162\!& 81\!& 72\!& 81\!&162\!& 9 &125\!& 50\!& 89\!& 40\!& 89\!&
50\!&125 \\
0 & 8 \!&144\!& 72\!& 64\!& 72\!&144\!& 8 &200\!&208\!& 40\!& 64\!& 40\!&
208\!&200 \\
-2& 9 \!&162\!& 81\!& 72\!& 81\!&162\!& 9 &125\!& 50\!& 89\!& 40\!& 89\!&
50\!&125 \\
-4& 18\!&324\!&162\!&144\!&162\!&324\!& 18& 26\!& 68\!& 50\!&208\!& 50\!&
68\!& 26 \\
-6& 1 \!& 18\!& 9 \!& 8 \!& 9 \!& 18\!& 1 & 1 \!& 26\!&125\!&200\!&125\!&
26\!& 1  \\
\hline
\end{tabular}
\end{equation*}\label{label101}
\end{table}

This exercise makes clear that the two QWs are radically different from
the outset. For the unentangled case, it is easy to see the result one may
already suspect from Fig.~\ref{fig:four}, that
the probability distribution of the unentangled 2D QW, $P(x,y) = P(x)P(y)$,
is a direct product of two 1D QWs in the $x$ and $y$ directions. We will
shortly demonstrate it numerically for large values of $N$. We remind the
reader that this result is not exactly intuitive, given that the walk steps
are each made in the $x$$\pm$$y$ direction of the lattice and as such would
appear to entangle the two directions completely; however, this result also
emerges from the 2D classical random walk (Sec.~IVA). For the entangled
case, the probability distribution has no direct relation to the 1D QW and
indeed already shows the concentration of probability along the $x$ and $y$
directions rather than along the diagonals.

\subsection{\label{sec5a} Cross-Sections of the Probability Distribution}

\begin{figure}[t]
\includegraphics[width=0.25\textwidth]{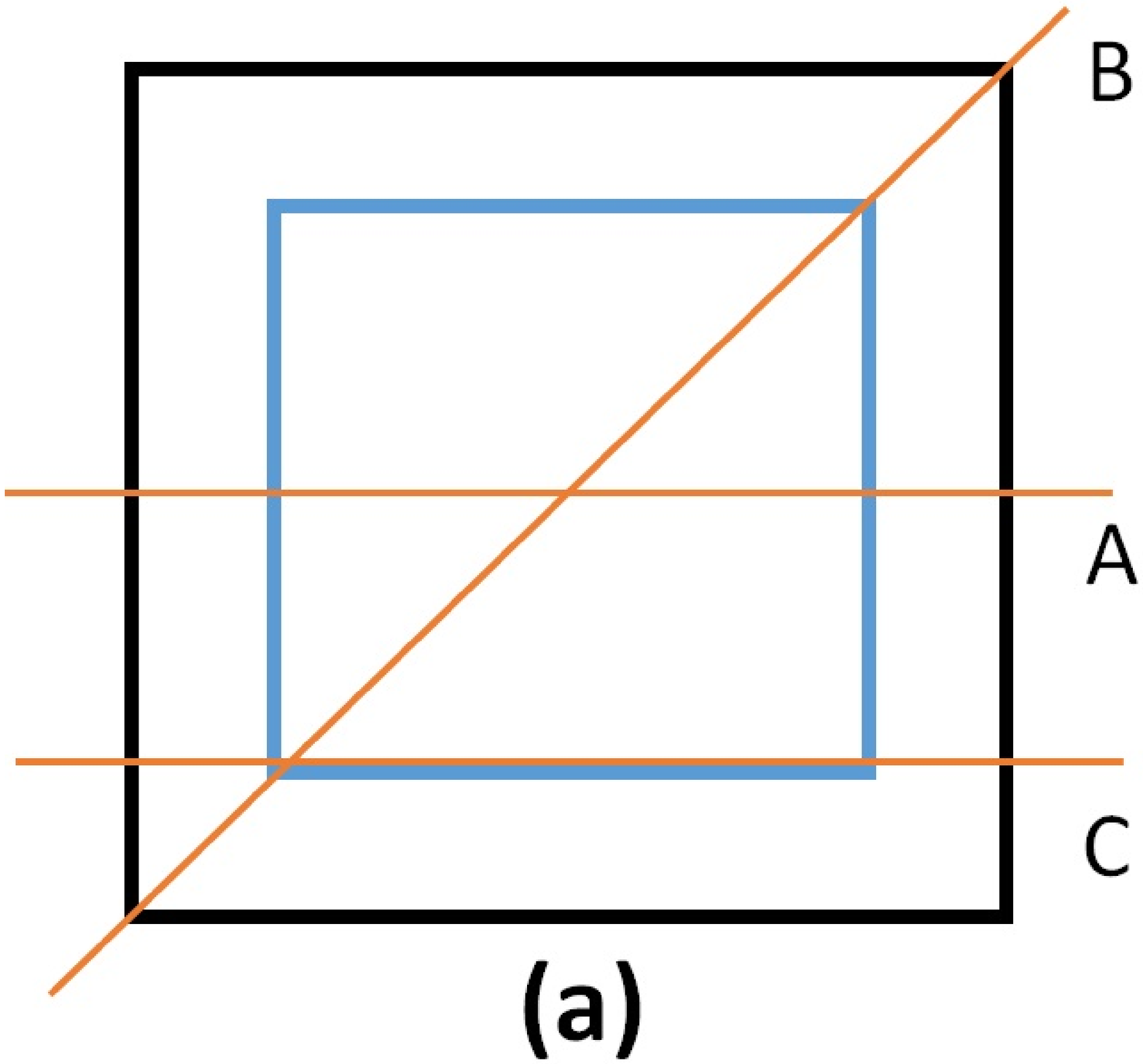}\includegraphics[width=0.25\textwidth]{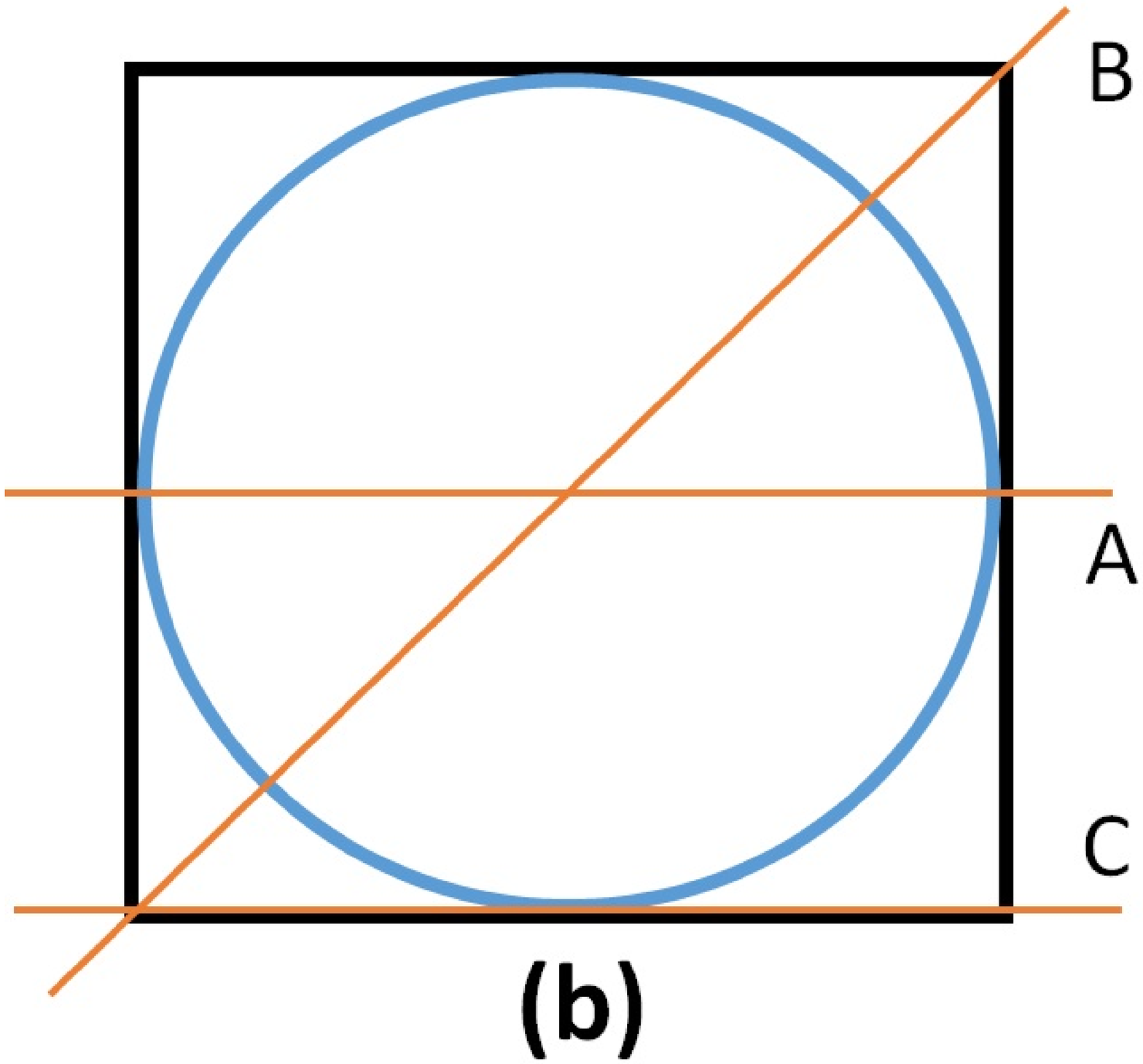}
\caption{Schematic illustration of 1D slices taken through the 2D
probability distribution for the two 2D QWs considered, whose data are
shown in Fig.~\ref{compare00}. Slice A denotes the $x$-direction, B the
diagonal, and C is the edge of data. In panel (a), where the blue square
denotes the region of maximum probability, slice C is taken at $x = 0.7 N$;
in panel (b), where the maximum probability is found along the blue circle,
slice C is taken at the true edge ($x = N$).} \label{compare0}
\end{figure}

For our numerical calculation of the characteristics of the different 2D
QWs, we have computed the probability distributions for both walks up to
$N = 1000$ steps, whose illustration requires a 1000$\times$1000 grid. To
show the results in a manner more quantitative than is possible in
Figs.~\ref{fig:four} and \ref{fig:2dqrw2cc}, we consider slices through the
2D probability data taken along the $x$ axis, along the diagonal $x$$+$$y$,
and along the ``edge'' of the data set, as shown in Fig.~\ref{compare0}.

\begin{figure}[t]
\includegraphics[width=0.5\textwidth]{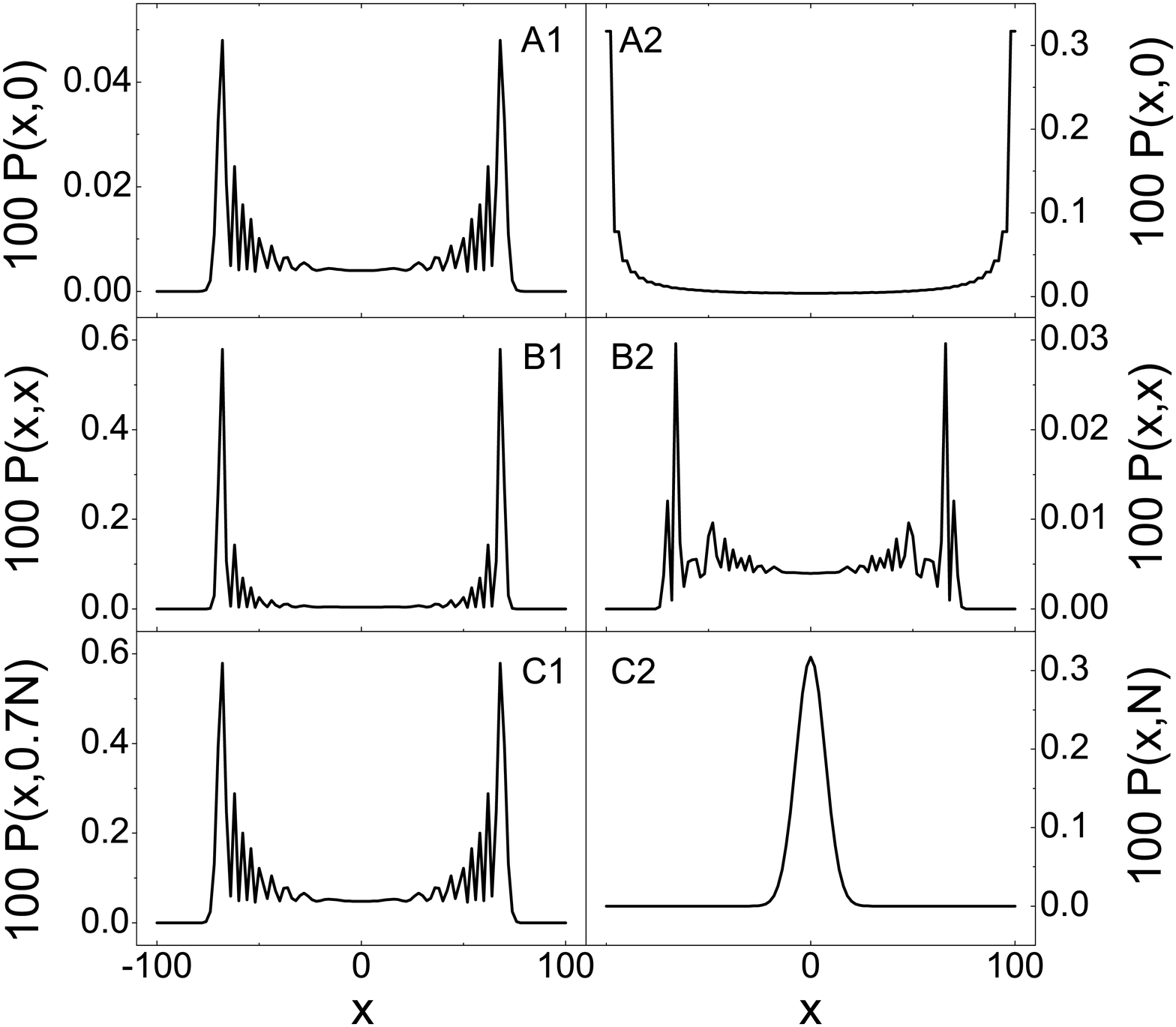}
\caption{Probability distributions on 1D slices through the 2D QW data
(see Fig.~\ref{compare0}) for walks of $N = 100$. Panels A1, B1, and C1
are for the unentangled QW and A2, B2, and C2 for the AQW.}
\label{compare00}
\end{figure}

The results of this process are illustrated in Fig.~\ref{compare00} for walks
of $N = 100$ steps. For the non-entangled case, panel A1 proves the numerical
identity with the 1D QW, $P(x,0) = P(x)P(0)$, which can be compared with
Fig.~\ref{fig:1dqrw} using the result for $P(0)$ with $N = 100$. This being
the case, panel C1 is very easy to interpret and for any parallel cut would
have the same functional form with any prefactor from $P(x)$. Panel B1 can
be expected to satisfy $P(x,x) = P^2(x)$, and therefore to have an envelope
function of the form $(N/\sqrt{2} - x)^{-1}$ across the outer half of the
distribution, a result we demonstrate below. By contrast, rather little is
known about the distribution $P(x,y)$ for the AQW and its understanding will
require applying the techniques of Sec.~III, for small data sets, to the
panels A2, B2, and C2 of Fig.~\ref{compare0}. Qualitatively, in A2, the
horizontal slice through the probability maxima, we observe an apparent
envelope function with no oscillations and a divergence towards $x = \pm N$
with an unknown power. In B2 we observe a complex oscillatory pattern with
significant probability extending beyond its peak value. In C2 we observe a
remarkably classical-looking probability distribution at the edge of the
system, where we remind the reader that the peak is the absolute maximum
anywhere in the walk.

\begin{figure}[t]
\includegraphics[width=0.5\textwidth]{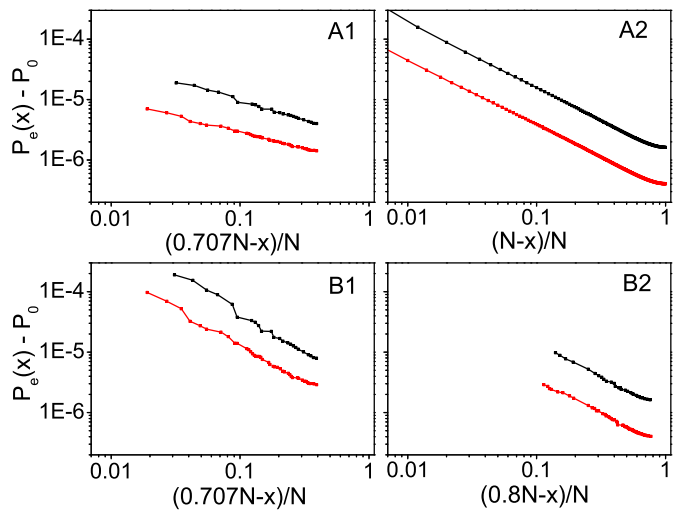}
\caption{(color online) Scaling properties of the envelope function of the
2D probability distribution for $N = 500$ (black) and $N = 1000$ (red) on
the data slices shown in Fig.~\ref{compare00}, panels A1, B1, A2, and B2. }
\label{compare001}
\end{figure}

We wish to characterize the AQW by the power-law form of its probability
envelope function around the peak values of the distribution. For this we
analyze the 2D probability slices following Eq.~(\ref{01e}), but because
our largest system size in 2D is $N = 1000$, the numerical results do not
give particularly reliable curve fits. Although the divergence of $P(x)$
at $x = N/\sqrt{2}$ (at $x = N$ for slice A2) is not achieved with any
accuracy for these values of $N$, we enforce this value in all cases other
than slice B2 to reduce the arbitrariness in the fitting parameters. We
show in Fig.~\ref{compare001} the probability on logarithmic axes as a
function of $(b-x)/N$ for fixed values $b = 0.707N$ in cases A1 and B1,
$b = N$ in case A2, and the fitted value $b = 0.8N$ in case B2. As expected,
the probabilities for panels A1 and B1 have gradients close to $-1/2$ and $-1$
[$c = 1/2$ and $c = 1$ in Eq.~(\ref{01e})], respectively, over the bulk of the
range, and the accuracy to which the data fall on a straight line benchmarks
the method for $N = 1000$. For panel A2, the data are remarkably clean and
give a clear gradient parameter $c = 1$, indicating that the envelope function
of the peak in the maximally entangled walk satisifes the form $P(x,0) = P_0
 + a/(1 - x/N)$ to high accuracy. For panel B2, a slice that does not include
the main peaks but only the circular edge of local probability maxima, we find
a result close to $c = 1/2$, although here the envelope is poorly defined and
the errors significant.

\begin{figure}[t]
\includegraphics[width=0.48\textwidth]{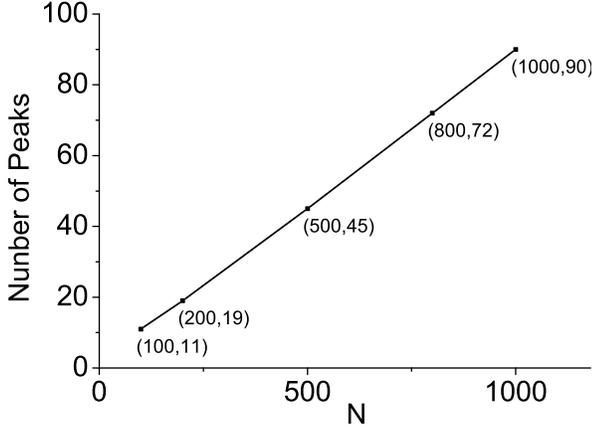}
\caption{Number of peaks in the probability distribution of the AQW (maximally
entangled two-coin QW) in the direction $x$$+$$y$, shown in panel B2 of
Fig.~\ref{compare00}, as a function of step number $N$. }
\label{cuttingb}
\end{figure}

While our estimate of the functional form of the $x$-dependence of
the AQW is somewhat approximate, the $N$-dependence of the probability
distribution is beyond doubt. At every corresponding position, the probability
falls by a factor of 4 for every doubling of $N$. This result is illustrated
by the black ($N = 500$) and red ($N = 1000$) lines in Fig.~\ref{compare001}.
Thus we propose that the appropriate fit to the probability data on these
four slices is given by the algebraic form
\begin{equation}
y = \frac{a_1}{N^2} + \frac{a_2}{N^{d}} (b - x)^{-c},
\label{2dfunc}
\end{equation}
with $c + d = 2$ and the values of $c$ as deduced from Fig.~\ref{compare001}.
We may conclude that the entangled 2D QW has algebraic scaling properties
similar to the 1D-case. While its probability distribution shows no
oscillations in the $x$ and $y$ directions, in the $x$$\pm$$y$ directions
it oscillates in a manner not dissimilar to the 1D QW (panel B2 in
Fig.~\ref{compare00}). In Fig.~\ref{cuttingb}, we show that the number of
peaks in this slice is again proportional to the step number $N$, and with
a constant of proportionality again of order 8.5\% (Sec.~IIIB), although we
point out that for $x \sim N/\sqrt{2}$, where the 1D QW shows its strongest
peak, this distribution has its deepest valley.

\subsection{\label{sec5b}Edge of the Probability Distribution}

Next we consider the probability distribution at the edge of the walk for
the AQW, shown in panel C2 of Fig.~\ref{compare00}. As remarked above,
this QW has the highly anomalous feature that its maximum probability
occurs when the walker takes the maximum number of steps in the same
direction along one of the two axes, but returns to the center of the
other axis. Further, a visual inspection of the 1D slice through this
maximum suggests that the distribution on this edge may be a Gaussian,
or a related function similar to the 1D classical random walk.

To investigate whether a quantum walk can lead to a classical results, we
employ the decomposition of the unitary evolution matrix into the matrices
$P$ and $Q$ introduced in Sec.~\ref{sec2b}. For the 2D QW we require
matrices $P_x$ for steps to the left, $Q_x$ for right, $P_y$ for up, and
$Q_y$ for down. Exploiting the AQW equivalence of the maximally entangled
two-coin QW, we use the same coin alternately for the $x$ and $y$
directions. Because our aim is to understand the probability distribution
on the right-hand edge of the system (equivalent to Fig.~\ref{compare0}(b),
slice C), every $x$-direction operation for the walker must be a $Q_x$ matrix
and not $P_x$. A complete cycle of the walk may then be either $R = P_y Q_x$
or $S = Q_y Q_x$, where
\begin{equation}
\begin{split}
& R = P_yQ_x = \frac{1}{2} \left( \!\! \begin{array}{cc} 1 & 1 \\ 0 & 0
\end{array} \!\! \right) \left( \!\! \begin{array}{cc} 0 & 0 \\ 1 & -1
\end{array} \!\! \right) = \frac{1}{2} \left( \!\! \begin{array}{cc}
1 & -1 \\ 0 & 0 \end{array} \!\! \right) \! , \\
& S = Q_y Q_x = \frac{1}{2} \left( \!\! \begin{array}{cc} 0 & 0 \\ -1 & 1
\end{array} \!\! \right) \left( \!\! \begin{array}{cc} 0 & 0 \\ 1 & -1
\end{array} \!\! \right) = \frac{1}{2} \left( \!\! \begin{array}{cc}
0 & 0 \\ 1 & -1 \end{array} \!\! \right) \! ,
\end{split}
\end{equation}
and thus
\begin{equation}
\begin{split}
& R R = \frac{1}{4} \left( \!\! \begin{array}{cc} 1 & -1 \\ 0 & 0 \end{array}
\!\! \right) \left( \!\! \begin{array}{cc} 1 & -1 \\ 0 & 0 \end{array} \!\!
\right) = \frac{1}{4} \left( \!\! \begin{array}{cc} 1 & -1 \\ 0 & 0 \end{array}
\!\! \right) = \frac{1}{2} R, \\
& R S = \frac{1}{4} \left( \!\! \begin{array}{cc} 1 & -1 \\ 0 & 0 \end{array}
\!\! \right) \left( \!\! \begin{array}{cc} 0 & 0 \\ -1 & 1 \end{array} \!\!
\right) = \frac{1}{4} \left( \!\! \begin{array}{cc} 1 & -1 \\ 0 & 0 \end{array}
\!\! \right) = \frac{1}{2} R, \\
& S R = \frac{1}{4} \left( \!\! \begin{array}{cc} 0 & 0 \\ -1 & 1 \end{array}
\!\! \right) \left( \!\! \begin{array}{cc} 1 & -1 \\ 0 & 0 \end{array} \!\!
\right) = \frac{1}{4} \left( \!\! \begin{array}{cc} 0 & 0 \\ -1 & 1 \end{array}
\!\! \right) = \frac{1}{2} S, \\
& S S = \frac{1}{4} \left( \!\! \begin{array}{cc} 0 & 0 \\ -1 & 1 \end{array}
\!\! \right) \left( \!\! \begin{array}{cc} 0 & 0 \\ -1 & 1 \end{array} \!\!
\right) = \frac{1}{4} \left( \!\! \begin{array}{cc} 0 & 0 \\ -1 & 1 \end{array}
\!\! \right) = \frac{1}{2} S,
\end{split}
\label{p02}
\end{equation}
i.e.~only the left-most operator in the string of steps is important.

With this result it is possible to calculate the entire edge distribution
analytically for any value of $N$. We illustrate the process for a walk
of $N = 4$ steps.
The paths arriving at position $(N,0)$ are $RRSS$,
$RSSR$, $RSRS$, $SSRR$, $SRSR$, and $SRRS$, which are divided between
the coin states $|\uparrow \rangle$ and $|\downarrow \rangle$ according to
\begin{equation}
\begin{split}
& U_{|\uparrow \rangle}^{(N,0)} = RRSS + RSSR + RSRS = \frac{1}{16} \left(
\!\! \begin{array}{cc} 3 & -3 \\ 0 & 0 \end{array} \!\! \right) \! , \\
& U_{|\downarrow \rangle}^{(N,0)} = SSRR + SRSR + SRRS = \frac{1}{16} \left(
\!\! \begin{array}{cc} 3 & -3 \\ 0 & 0 \end{array} \!\! \right) \! ,
\end{split}\nonumber
\end{equation}
whence
\begin{equation}
P_{|\uparrow \rangle}^{(N,0)} = \left| \frac{3-3i}{16\sqrt{2}} \right| =
\frac{9}{256}, \;\; P_{|\downarrow \rangle}^{(N,0)} \; = \; \left| \frac{3-3i}
{16\sqrt{2}} \right| = \frac{9}{256}, \nonumber
\end{equation}
and finally
\begin{equation}
P^{(N,0)} = P_{|\uparrow\rangle}^{(N,0)} + P_{|\downarrow\rangle}^{(N,0)} =
\frac{18}{256}.
\label{epaqwe0}
\end{equation}
Paths arriving at position $(N,2)$ are $RRRS$, $RRSR$, $RSRR$, and $SRRR$,
yielding
\begin{equation}
\begin{split}
& U_{|\uparrow \rangle}^{(N,2)} = RRRS + RRSR + RSRR = \frac{1}{16} \left(
\!\! \begin{array}{cc} 3 & -3 \\ 0 & 0 \end{array} \!\! \right) \! , \\
 &U_{|\downarrow \rangle}^{(N,2)}= SRRR =\frac{1}{16} \left( \!\! \begin{array}{cc}
0 & 0 \\ -1 & 1 \end{array} \!\! \right), \\ & P_{|\uparrow \rangle}^{(N,2)} =
\left| \frac{3-3i}{16\sqrt{2}} \right| = \frac{9}{256}, \;\; P_{|\downarrow
\rangle}^{(N,2)} \; = \; \left| \frac{-1+i}{16\sqrt{2}} \right| = \frac{1}{256},
\end{split}\nonumber
\end{equation}
and thus
\begin{equation}
P^{(N,2)} = P_{|\uparrow \rangle}^{(N,2)} + P_{|\downarrow \rangle}^{(N,2)} =
\frac{10}{256}.
\label{epaqwe2}
\end{equation}
The only path arriving at position $(N,4)$ is $RRRR$, leading to
\begin{eqnarray}
U_{|\uparrow \rangle}^{(N,4)} & = & RRRR \; = \; \frac{1}{16} \left( \!\!
\begin{array}{cc} 1 & -1 \\ 0 & 0 \end{array} \!\! \right) \! , \nonumber \\
P^{(N,4)} & = & P_{|\uparrow \rangle}^{(N,4)} = \left| \frac{1-i}{16\sqrt{2}}
\right| = \frac{1}{256}.
\label{epaqwe4}
\end{eqnarray}
This analytical solution demonstrates that the quantum mechanical
interference leading to the probability on the edge of the system is
constructive everywhere, with none of the paths interfering destructively.
Regions at the center of the edge simply have the most paths, and this is
the origin of what we can call the ``semi-classical'' result that the
probability is maximal at the center of the edge, leading to a distribution
similar in appearance to the binomial distribution of the classical random
walk. In fact the degree of destructive interference everywhere else in the
2D AQW is such that these maxima on the edges are the global maxima, meaning
that the walker is effectively pushed to the maximal number of steps in order
that it does not ``destroy itself'' by interference in the maximally entangled
two-coin walk.

\begin{table}[b]
\caption{Probabilility tables for the first 7 steps of the binomial
distribution (top), omitting a prefactor of $1/2^N$ at each level $N$
of the table, and for the pseudobinomial distribution (bottom) achieved
on the edges of the 2D AQW, omitting a prefactor of $1/4^N$.}
\begin{equation*}
\begin{array}{cc}
\begin{tabular}{|ccccccccccccccc|}
\hline  & & & & & & & 1 & & & & & & &   \\
\hline  & & & & & & 1 & & 1 & & & & & &\\
\hline  & & & & & 1 & & 2 & & 1 & & & & &  \\
\hline  & & & & 1 & & 3 & & 3 & & 1 & & & &  \\
\hline  & & & 1 & & 4 \ & & 6 \ & & 4 \ & & 1 & & & \\
\hline  & & 1 & & 5 & & 10 & & 10 & & 5 & & 1 & & \\
\hline  & 1 & & 6 & & 15 & & 20 & & 15 & & 6& & 1 &\\
\hline  1 & & 7 & & 21 & & 35 & & 35 & & 21& & 7 & & 1 \\
\hline
\hline  & & & & & & & 1 & & & & & & &  \\
\hline  & & & & & & 1 & & 1 & & & & & & \\
\hline  & & & & & 1 & & 2 & & 1 & & & & &  \\
\hline  & & & & 1 & & 5 & & 5 & & 1 & & & &  \\
\hline  & & & 1 & & 10 & & 18 & & 10 & & 1 & & & \\
\hline  & & 1 & & 17 & & 52 & & 52 & & 17 & & 1 & & \\
\hline  & 1 & & 26 & & 125 & & 200 & & 125 & & 26& & 1 & \\
\hline  1 & & 37 & & 261 & & 625 & & 625 & & 261& & 37 & & 1 \\
\hline
\end{tabular}
\end{array}\end{equation*}\label{label203}
\end{table}

\begin{figure}[t]
\includegraphics[width=0.48\textwidth]{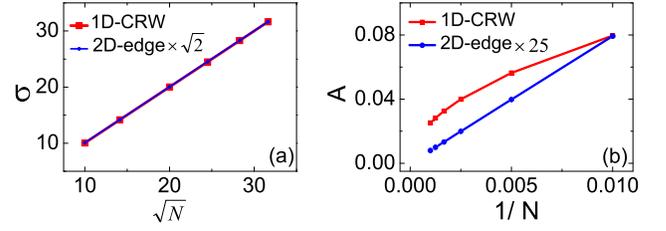}
\caption{Comparison between the AQW edge probability distribution and a
binomial (classical random walk) distribution for values up to $N = 1000$
steps, where the binomial is very accurately Gaussian. (a) Standard
deviation $\sigma$. (b) Prefactor $A$.} \label{fitcompare}
\end{figure}

A quantitative examination of the coefficients of the AQW edge reveals
that, despite being peaked at the center, they are not the same as the 1D
classical random walk, as shown in Table \ref{label203}. The probability
distribution at position $x$ on the edge of an $N$-step AQW can in fact be
expressed exactly as
\begin{eqnarray}
P_N (N,x) & = & {\textstyle {\frac{1}{4}}}[P_{N-1}^2 (x-1) + P_{N-1}^2 (x+1)]
\nonumber \\ & = & \frac{1}{4^N} \left( \left[ C_{N-1}^{\frac{N-x-1}{2}} \right]^2
 + \left[ C_{N-1}^{\frac{N-x+1}{2}} \right]^2 \right) \! ,
\label{eq02}
\end{eqnarray}
where $C_n^i$ is a binomial coefficient (Sec.~IIA). This remarkable result,
which we term a ``pseudobinomial'' distribution, encodes the emergence of
the physics of classical probabilities in the highly entangled QW. The
binomial coefficients appearing in the penultimate step before measurement
(Table \ref{label203}) arise as a consequence of the complete constructive
interference of paths, as illustrated in Eqs.~(\ref{epaqwe0}) to
(\ref{epaqwe4}) and discussed in the preceding paragraph.

To demonstrate the form of the AQW edge probability at large $N$, we fit both
binomial [Eq.~(\ref{eq01})] and pseudobinomial distributions to the Gaussian,
as specified in Eq.~(\ref{egf}) and again with $P_0 = 0 = b$, to compare
their forms and to extract their standard deviations $\sigma$. As shown in
Fig.~\ref{fitcompare}(a), while the binomial distribution gives the result
$\sigma = \sqrt{N}$, the standard deviation of the AQW edge probability is
$\sigma = \sqrt{N/2}$, i.e.~the distribution is narrower and the diffusion,
or spreading, rate of the walk slower (more localized) by a factor of
$1/\sqrt{2}$. Concerning the normalization prefactor $A$
[Fig.~\ref{fitcompare}(b)], the binomial approaches the Gaussian result
$A = 1/\sqrt{2 \pi N}$, but the AQW edge distribution does not, following
instead a dependence $A \propto 1/N$. While the probability of the classical
random walk of course sums to unity, determining $A$ for a true Gaussian,
the probability distribution at the edge of the AQW is not normalized due
to the weight in the interior of the walk, and hence the edge probability
is found to decay more rapidly than a true Gaussian. The emergence of such
a pseudobinomial distribution in a QW is yet another example of the rich
physics contained in a deceptively simple quantum evolution algorithm.

\begin{figure}[t]
\includegraphics[width=0.5\textwidth]{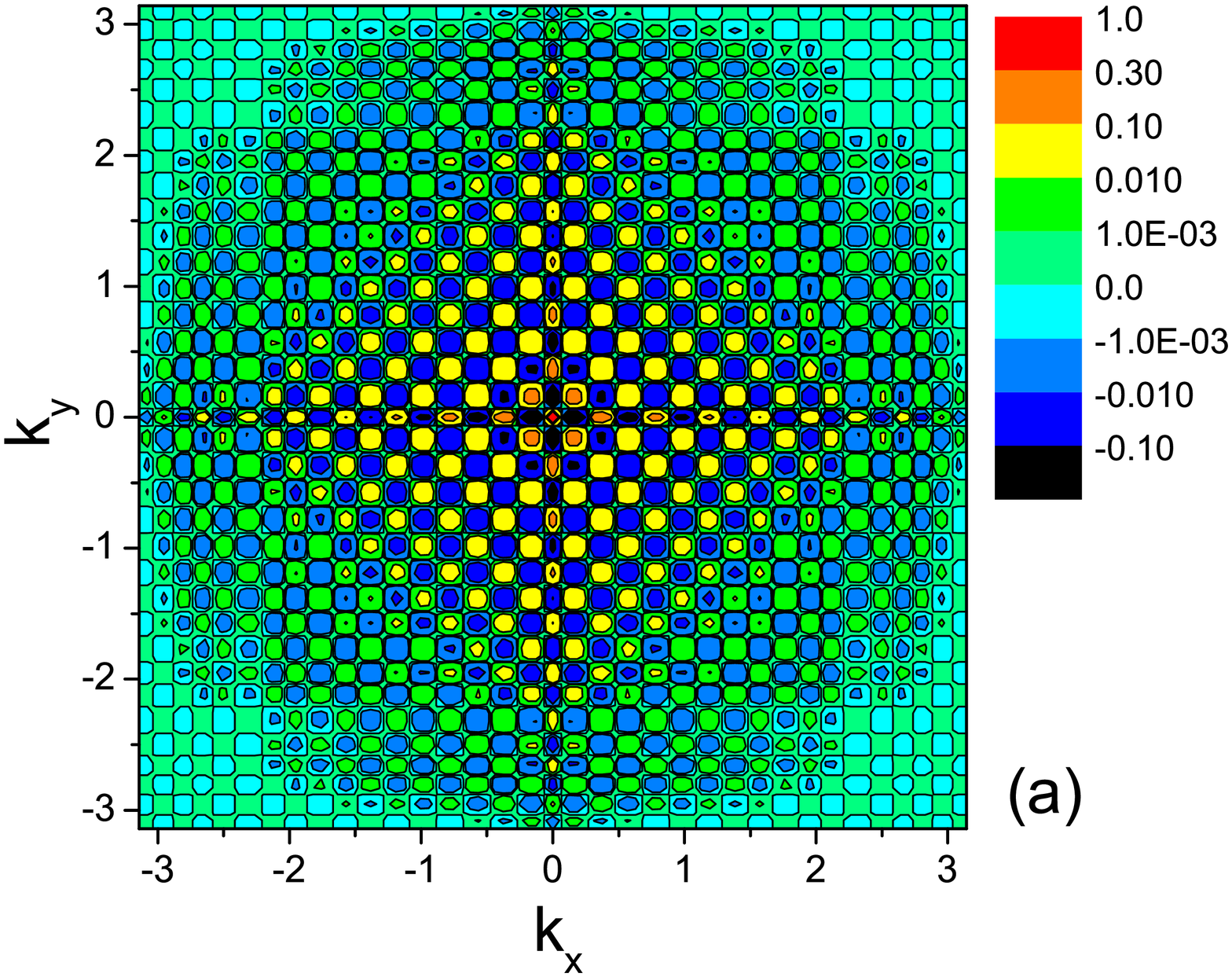}
\includegraphics[width=0.5\textwidth]{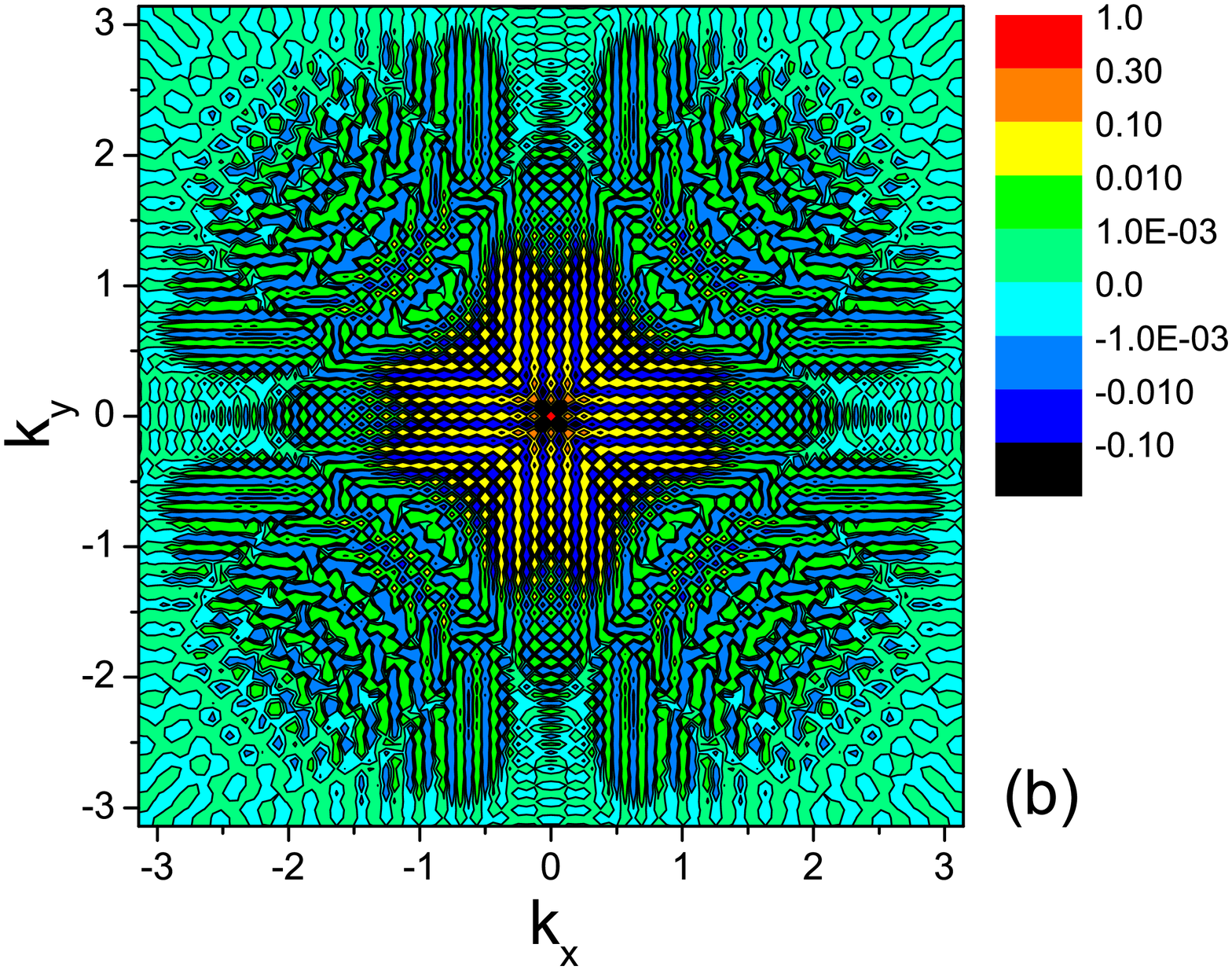}
\caption{2D Fourier transform $F(k_x,k_y)$ of the probability distributions
$P(x,y)$ (Fig.~\ref{fig:2dqrwc}) for (a) the four-face-coin or unentangled
2D QW and (b) the 2D AQW or maximally entangled two-coin QW, both for
$N = 100$.}
\label{ft2d}
\end{figure}

\subsection{\label{sec5c} Fourier Transformation of 2D QWs}

For further insight into the structure of the AQW probability distribution,
we compute the Fourier transform $F(k_x,k_y)$ of $P(x,y)$ for both the 2D
QWs we consider. Figure \ref{ft2d}(a) shows the 2D Fourier transform of
the unentangled (four-face-coin) QW, whose product structure is again
clearly visible. It was shown in Sec.~III that the 1D QW has spatial
frequency information at all scales up to the inverse step size ($k = \pi$)
and this is clear in the finite components of $F(k_x,k_y)$ up to the edge
of Fourier space. The AQW, shown in Fig.~\ref{ft2d}(b), is again quite
different, showing both an intrinsically 2D character and an apparent
concentration of weight closer to the center of Fourier space.

\begin{figure}[t]
\includegraphics[width=0.5\textwidth]{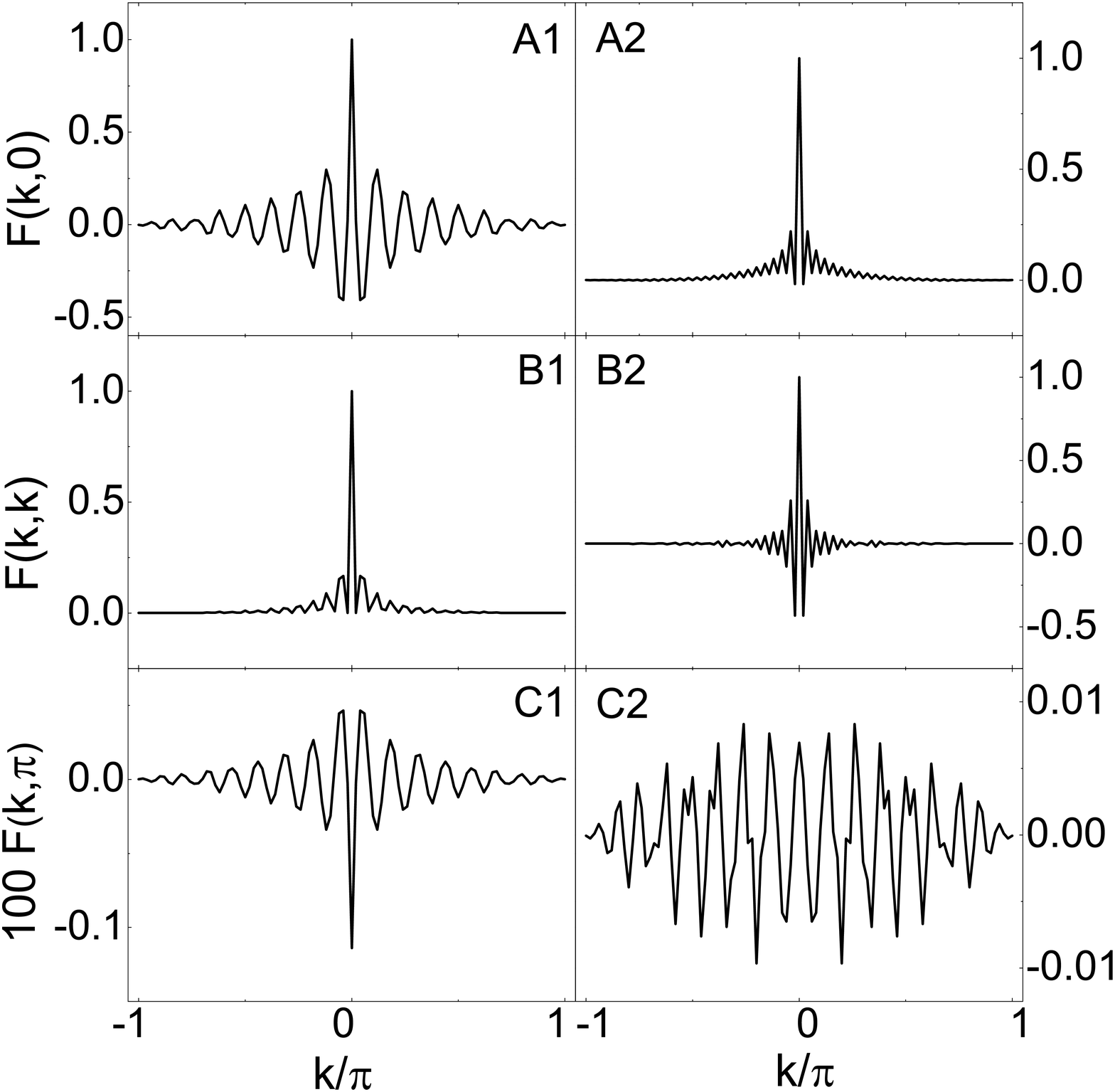}
\caption{Cross-sections of the 2D Fourier transform $F(k_x,k_y)$ of the
probability distributions [Fig.~\ref{compare00}] for the two 2D QWs of
Sec.~IV, calculated with $N = 100$ and shown for the slices studied in
Sec.~VA (Fig.~\ref{compare0}).}
\label{FTcutting1}
\end{figure}

To analyze these distributions in a quantitative manner, we present the
data in the form of 1D slices. Figure \ref{FTcutting1} is completely
analogous to Fig.~\ref{compare00}, with the unentangled QW in the left
panels, the entangled one on the right, A denoting a horizontal slice
through the center of the Fourier distribution, B a diagonal slice, and
C the edge. We illustrate the qualitative features of the data using QWs
of $N = 100$ steps, but for our numerical analysis of the properties of
the Fourier transforms we use values of $N$ up to 1000.

In panel A1 of Fig.~\ref{FTcutting1}, which shows $F(k_x,0)$ for the
unentangled QW, we observe the factorized form $F(k_x)F(0) = F(k_x)$
expected from Sec.~VA and shown in Fig.~\ref{num-ana}(b). Panel B1 shows
$F(k_x,k_x) = F^2(k_x)$, with only positive components and a correspondingly
steeper decay of the envelope. Panel C1, showing $F(\pi,k_y)$, is identical
to A1 up to a multiplicative prefactor, which is small and happens to be
negative at $k_x = \pi$. Turning to the AQW, where the probability
distribution cannot be factorized, the slice $F(k_x,0)$ in panel A2 is
not dissimilar to panel B1, in that all of the Fourier components are
positive and they decay significantly more rapidly than those of the 1D
QW. However, their oscillatory form shows a very precise odd/even
modulation, which is not evident in the 1D QW. The diagonal slice
$F(k_x,k_x)$ is shown in panel B2 and confirms both the odd/even
modulation and the concentration of Fourier amplitude near the center
of the system. Finally, panel C2 has no readily discernible structure,
reflecting the fact that the AQW edge is non-oscillatory and thus
dramatically different from all the other QW distribution slices (Sec.~VB).

\begin{figure}[t]
\includegraphics[width=0.48\textwidth]{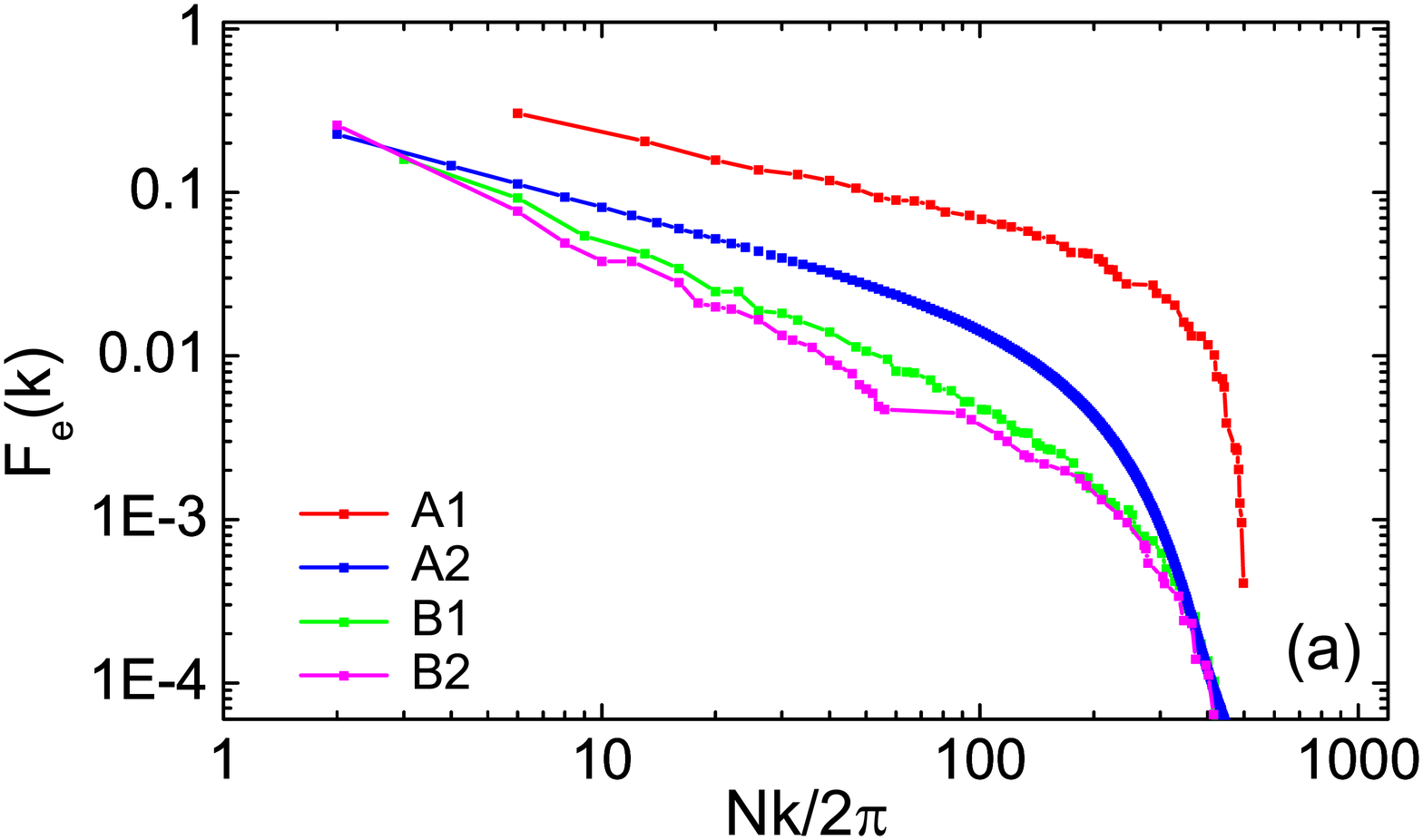}
\includegraphics[width=0.48\textwidth]{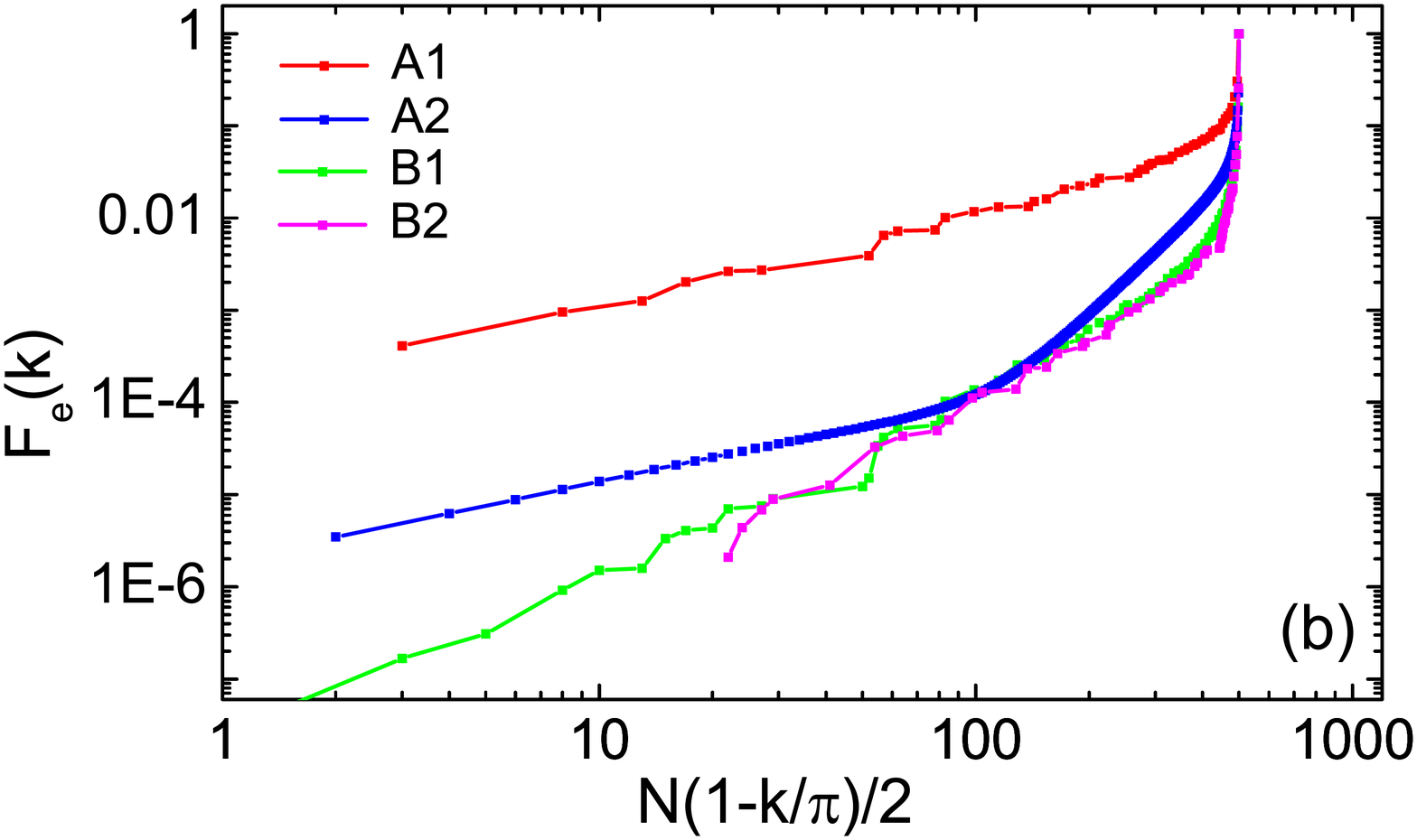}
\caption{Fourier transform of the 2D probability distributions for slices
A1, B1, A2, and B2 shown in Fig.~\protect{\ref{compare00}}, (a) as a function
of $k$, showing algebraic decay of the small-$k$ components, and (b) as a
function of $1 - k/\pi$, showing different power-law dependences at large $k$.}
\label{FTcutting2}
\end{figure}

Following Sec.~IIID, we investigate two properties of the Fourier
transform slices, namely their envelope and their oscillations. To
characterize the decay of the envelope function, we follow the procedure
shown in Fig.~\ref{fit02} and use logarithmic axes. As shown in
Fig.~\ref{FTcutting2}(a), and as expected from Sec.~IIID, the envelope
functions for panels A1 and B1 yield algebraic decay exponents close to
$c = 1/2$ and $c = 1$, respectively, for the Fourier components around
$k = 0$; the accuracy with which the data adhere to a straight line again
benchmarks the accuracy one may expect from data up to $N = 1000$. The data
from slice A2 form by far the best-quality set in Fig.~\ref{FTcutting2}(a)
and the algebraic decay exponent is unequivocally $c = 1/2$, meaning that
the spatial periodicity content of the AQW is qualitatively similar to
that of the unentangled QW at low frequencies. We take this opportunity
to remind the reader that there is {\it a priori} no direct connection
between the real- and reciprocal-space exponents of the envelope functions
(respectively around the peak of $P(x)$ and around $k = 0$) and the two
QWs present an example pair with different exponents in real space,
describing the peak shape, but the same exponent in reciprocal space,
describing the frequency content. The data from slice B2 exhibit the
lowest-quality envelope in the figure, but still show a strong qualititive
similarity to slice B1, with $c = 1$. Again this highlights the complexity
of the spatial frequency content of the AQW in its different directions
and indicates that the entanglement inherent in the AQW also entangles the
spatial information of the different lattice directions.

For completeness, we show in Fig.~\ref{FTcutting2}(b) the Fourier components
of the four slices at high $k$, where once again slices A1 and B1, with
respective gradients $c' = 1$ and $c' = 2$ benchmark the accuracy of the
approach. Again we observe that slices A2 and B2 for the AQW are qualitatively
similar, with the caveat that the data for slice A2 show evidence of a
distinctively different intermediate regime. We stress that the difference
in amplitude of the Fourier components between the unentangled QW and the
AQW, which reaches two orders of magnitude over the large-$k$ half of the
range for slices A1 and A2, is the most meaningful measure of strong
quantitative differences between the two QWs.

Concerning the oscillations visible in the Fourier transforms, in the
unentangled case it is known from Sec.~III that the 2D QW contains
spatial frequency information on all length scales and that beating
phenomena become visible both in real space around the maximum frequency
and as a result in Fourier space near $k = \pm \pi$. For the AQW, our data
(Figs.~\ref{FTcutting1} and \ref{FTcutting2}) show that the high-frequency
spatial components are very significantly weaker, meaning that lattice-scale
oscillations are of little relevance, and this suggests that beating (which
we are unable to find up to $N = 1000$) is unlikely to be present. We believe
that the extra strength of even harmonics in the AQW results from the even
step number.

We close our analysis of the Fourier-space information contained in a QW
by summarizing a complementary perspective, named the ``dispersion relation''
approach in Ref.~\cite{PhysRevA.87.022336}. In Sec.~IIC we presented the
exact solution of the 1D QW by considering its nature in Fourier space,
finding that the Hadamard operator (\ref{eho}) gave rise to a dispersion
relation for a quantity $\omega_k = \sin^{-1} [\sin k/\sqrt{2}]$, specified
in terms of $k$. Although $\omega_k$ is largely a simplifying notation, as
there is no concept of a separate ``time'' (step number) and space in a
QW, it does result in a compact description in higher dimensions and it
has the added advantage of reflecting different degrees of coin entanglement
in a transparent way. When the AQW in 2D is transformed into Fourier space
\cite{PhysRevA.87.022336}, one may express the operator as
\begin{equation}
G_4 = \frac{1}{2} \! \left( \! \begin{array}{cccc}
\! - e^{ik_1} e^{ik_2} \! & \!\! e^{ik_1} e^{ik_2} \! & \!\! e^{ik_1} e^{ik_2} \! &
\!\! e^{ik_1} e^{ik_2} \! \\ \! e^{ik_1} e^{-ik_2} \! & \!\! - e^{ik_1} e^{-ik_2} \! &
\!\! e^{ik_1} e^{-ik_2} \! & \!\! e^{ik_1} e^{-ik_2} \! \\ \! e^{-ik_1} e^{ik_2} \! &
\!\! e^{-ik_1} e^{ik_2} \! & \!\! - e^{-ik_1} e^{ik_2} \! & \!\! e^{-ik_1} e^{ik_2} \!
\\ \! e^{-ik_1} e^{-ik_2} \! & \!\! e^{-ik_1} e^{-ik_2} \! & \!\! e^{-ik_1} e^{-ik_2} \!
 & \!\! - e^{-ik_1}e^{-ik_2} \! \end{array} \!\!\! \right) \nonumber
\label{eg4}
\end{equation}
and deduce the eigenvalues
\begin{equation}
\begin{aligned}
\lambda_k^{1\pm} & = \pm 1, \\
\lambda_k^{2\pm} & = - {\textstyle \frac{1}{2}} [\cos(k_1 + k_2) + \cos(k_1 -
k_2)  \\ & \;\; \pm \sqrt{(\cos(k_1 + k_2) + \cos(k_1 - k_2))^2 - 4} ].
\end{aligned}
\label{ee01}
\end{equation}
Because $\lambda_k = e^{i\omega_k}$, the dispersion relations for the four
eigenmodes of the AQW take the form
\begin{equation}
\begin{aligned}
\omega_k^{1+} & = 0, \;\;\;\; \omega_k^{1-} = \pi, \\
\omega_k^{2\pm} & = \pi \mp {\textstyle \frac{1}{2}} (\cos k_1 + \cos k_2),
\end{aligned}
\label{ee02}
\end{equation}
allowing a considerable simplification of the process [Eq.~(\ref{f01})] of
generating the probability distribution in this case.

\section{\label{sec7} Summary}

Quantum walks produce probability distributions entirely different
from the well-known classical ``drunkard's walk.'' In most cases, the
distribution is controlled completely by predominantly destructive
interference between the paths returning to the center, with the result
that the positions of maximal probability are pushed out towards the edges
of the walk. This results in the ``linear diffusion'' property of a quantum
walker, which has made the QW evolution algorithm so valuable in studies of
quantum computing.

However, as quantum computing approaches large-scale implementation, it is
necessary to understand the nature of quantum algorithms at large scales.
By this is meant an exact account of their properties, information content,
limiting behavior, and scaling characteristics at long evolution times.
Although analytical studies of different quantum walks in one and two
dimensions have revealed certain characteristics of their evolution and
interference, in particular their dependence on the initial state and the
entanglement of the quantum coins generating them, to date there has been
rather little consideration of the situation at large step number $N$.

By calculating the probability distributions for one- and two-dimensional
quantum walks up to $N = 1000000$ in the former case and $N = 1000$
in the latter, we have revealed a number of properties and scaling
characteristics. In one dimension, we verify that the probability approaches
peaks at $x/N = \pm \, 1/ \! \sqrt{2}$ at large $N$, which is the transition
region between destructive interference and vanishing probability for all
steps to be made in the same direction. The normalized support converges to
the region $[- 1/ \! \sqrt{2},1/ \! \sqrt{2}]$ and the envelope of the
distribution peaks has a square-root decay, i.e.~an algebraic form. Within
this envelope, the probability shows systematic oscillations on all length
scales, with the number of probability peaks always the same fraction of $N$.
The frequency of these oscillations increases from the inverse system size to
the inverse step length as a function of distance from the center of the walk.
The different frequencies show complex beating phenomena in the regime where
the oscillations are most rapid. These properties are revealed in a
complementary fashion by taking the spatial Fourier transform of the
distribution. All of these features are universal for walks of all $N$
values, giving them very strong similarities, but not the property of
self-similarity (there are no fractal structures in the simple walks
studied here).

In two dimensions there is not one quantum walk, or even one
algorithm, but a spectrum of protocols capable of generating quantum
evolution in a plane of phase space. We study two examples
that in fact represent the limiting cases of the entanglement between
two orthogonal directions: unentangled 2D QW and the maximally entangled alternate quantum walk(AQW).

In our numerical studies of these two limits, the understanding developed
in one dimension gives a complete account of the unentangled quantum walk,
whose two-dimensional probability distribution factorizes exactly into two
one-dimensional functions. By contrast, the maximally entangled case exhibits
strong correlations between the two orthogonal directions, damping of the
oscillatory behavior, and the extraordinary feature that the maximum of the
probability distribution is pushed all the way to the system edge by the
dominance of destructive interference. We provide an analytical description
of the edge distribution, showing that all paths arriving at the system
edges interfere constructively and proving that its functional form is a
type of pseudo-binomial, which is semi-classical in the sense of approaching
a Gaussian dependence on the spatial coordinate at large $N$.

As two-dimensional quantum walks become an experimental science,
our analytical and numerical studies demonstrate that even the simplest
algorithms for quantum evolution contain a rich variety of physical
phenomena and potential for technological application.

\acknowledgments

The authors gratefully acknowledge helpful discussions with Professor B.Normand.
Work at Renmin University was supported by the National Natural Science Foundation
of China (NSFC) under Grant No.~11174365 and by the National Basic Research
Program of China (NBRPC) under Grant No.~2012CB921704. PX was supported by
the NSFC under Grant Nos.~11174052 and 11474049, by the NBRPC under Grant
No.~2011CB921203, by the Open Fund from the State Key Laboratory of Precision
Spectroscopy of East China Normal University, and by the CAST Innovation Fund.

\bibliographystyle{apsrev4-1}
\bibliography{a}

\end{document}